\documentclass[aps,pra,floatfix,twocolumn,superscriptaddress]{revtex4-1}

\usepackage{mathrsfs}
\usepackage{graphicx}
\usepackage[english]{babel}
\usepackage{amsmath}
\usepackage{amssymb}

\usepackage{relsize}
\usepackage{dsfont}
\usepackage{bm}% bold math

\newcommand{\me}{\mathrm{e}}
\newcommand{\mi}{\mathrm{i}}

\newcommand{\dif}{\mathrm{d}}

\newcommand{\bR}{\mathbf{R}}
\allowdisplaybreaks
\begin{document}

\title{Dynamic process and Uhlmann process: Incompatibility and dynamic phase of mixed quantum states}
\author{Hao Guo}
\affiliation{Department of Physics, Southeast University, Jiulonghu Campus, Nanjing 211189, China}

\author{Xu-Yang Hou}
\affiliation{Department of Physics, Southeast University, Jiulonghu Campus, Nanjing 211189, China}
\author{Yan He}
\affiliation{School of physics, Sichuan University, Chengdu, Sichuan 610064, China}
\author{Chih-Chun Chien}
\email{cchien5@ucmerced.edu}
\affiliation{Department of physics, University of California, Merced, CA 95343, USA}

\begin{abstract}
While a pure quantum state may accumulate both the Berry phase and dynamic phase as it undergoes a cyclic path in the parameter space, the situation is more complicated when mixed quantum states are considered. From the Ulhmann bundle, a mixed quantum state can accumulate the Ulhmann phase if the parallel-transport condition is satisfied. However, we show that the Ulhmann process is in general not compatible with the evolution equation of the density matrix governed by the Hamiltonian. Thus, a mixed quantum state usually accumulates a dynamic phase during its time evolution. We present the expression of the dynamic phase for mixed quantum states. In examples of one-dimensional two-band models and simple harmonic oscillator, the dynamic phase can take multiple discrete values in quasi-static processes at infinitely high temperature due to the  resonant points. However, the behavior differs if the energy spectrum is continuous without a band gap. Moreover, there is no natural analog of the dynamic phase in classical systems.
\end{abstract}

\maketitle

%\tableofcontents
%\newpage

\section{Introduction}
The geometric phases have played an important role in many quantum phenomena ever since their modern applications in quantum systems~\cite{Berry84,Bohm03,Vanderbilt_book,Cohen19}. By using the electromagnetic duality~\cite{Franson99}, there can be various ways for inducing the geometric phase. The Berry phase also offers the first step towards the understanding of topological insulators and superconductors~\cite{TKNN,Haldane,KaneRMP,ZhangSCRMP,MooreN,KaneMele,KaneMele2,BernevigPRL,MoorePRB,FuLPRL,Bernevigbook,ChiuRMP}. However, the Berry-phase formalism has been developed for pure quantum states. A fundamental question is what is the generalization of the Berry phase from pure quantum states to mixed quantum states?
Uhlmann mathematically formulated a possible extension of the Berry phase to mixed quantum states ~\cite{Uhlmann86,Uhlmann89,Uhlmann91,Uhlmann96}. Interestingly, the generalization is made by a mathematician instead of a physicist. Hence, the early discussion of the Uhlmann phase was quite technical.

More recently, the geometry and topology of mixed quantum states has been explored~\cite{GPbook,HZUPPRL14,DiehlPRB15}. The Uhlmann phase, which was proposed to play the role of the Berry phase for mixed states, has been calculated for several topological systems in Refs.~\cite{ViyuelaPRL14,ViyuelaPRL14-2,2DMat15}. While experimental measurements of the Uhlmann phase of selected systems have been reported in Ref.~\cite{npj18}, other possible means for interpreting or measuring geometric phases for mixed states have been proposed~\cite{GPMQS1,Marzlin04}. There have been attempts to reconcile the topological
criteria for mixed  states~\cite{GPMQS1,GPMQS2,GPMQS3,GPMQS4,GPMQS5,GPMQS6,DiehlPRB15,DiehlPRX17,Uhlmann17,MeraPRL17,Uhlmann18,UPPRB18,HYUPPRB18,UPPRA18}. It was found that the Uhlmann bundle for constructing the Uhlmann phase is trivial~\cite{DiehlPRB15}, and a full understanding of the topological properties of mixed quantum states still awaits future research~\cite{Asorey19}.

The Berry phase is produced when a pure quantum state undergoes an adiabatic process, and the corresponding theory can be expressed in an elegant geometrical language by introducing the parallel-transport condition of a quantum state (or a fibre element in mathematics)~\cite{Nakahara}. We will briefly summarize the geometric formalism of the Berry phase. Importantly, the parallel-transport condition is compatible with the dynamic evolution of the system according to its Hamiltonian. As a consequence, one can write down a combined "Berry process" for describing how a system can acquire both the Berry phase and the dynamic phase during its time-evolution.

Uhlmann has generalized the parallel-transport condition from the fiber bundle of pure quantum states to that of mixed quantum states expressed by the density matrices~\cite{Uhlmann86,Uhlmann89,Uhlmann91,Uhlmann96}. However, it is not clear what physical implication corresponds to the parallel transport of the quantum density matrices.
Although the Uhlmann phase was constructed as a generalization of the Berry phase, it may not be necessarily related to the adiabatic evolution of a mixed quantum states. Some literatures instead regard the Uhlmann phase as a generalization of the Aharonov-Anandan phase~\cite{GPbook,Asorey19}.

While the derivation of the Uhlmann phase could be performed with the aid of the Bures distance between density matrices~\cite{Hubner93}, here we show that it can be done by a unified derivation from a geometric formalism using Ref.~\cite{Nakahara}. A more mathematical treatment can also be found in Ref.~\cite{Asorey19}.
Importantly, the unified derivation allows us to identify a fundamental difference between the Berry process and the Uhlmann process, defined by parallel-transport in their corresponding fiber bundles. A quantum system is evolved according to its Hamiltonian~\cite{MessiahBook,Merzbacher_book}, regardless if it is in a pure or mixed quantum state. For the Berry phase, the parallel-transport condition is compatible with the dynamic process governed by the Hamiltonian if the adiabatic condition holds. Therefore, one can combine the parallel-transport with the dynamic process and define an effective "adiabatic Hamiltonian", which governs the dynamics and cause the system to acquire both the Berry phase and the dynamic phase.

In stark contrast, the parallel-transport condition of the Uhlmann bundle is incompatible with the dynamic process of the density matrix governed by the Hamiltonian, as we will show later. Therefore, a mixed quantum state violates the parallel-transport condition of the Uhlmann bundle when it evolves according to its Hamiltonian. The phase accumulated by the mixed state during the dynamic process is thus not the Uhlmann phase but the dynamic phase. While the dynamic phase of pure quantum states has been presented in textbooks~\cite{MessiahBook,Merzbacher_book}, the definition of the dynamic phase of mixed quantum states is not universal in the literature. Here we present a generalization of the dynamical phase from pure quantum states to mixed quantum states according to the procedure that leads to the Uhlmann phase, albeit the dynamic process, not the Uhlmann process, is followed.

The dynamic phase of two selected one-dimensional (1D) two-band model and the simple harmonic oscillator will be presented. Being not a geometric phase, the dynamic phase does not carry topological information. Nevertheless, we found resonant behavior that pins the value of the dynamic phase due to the energy gap or energy spacing. Moreover, the dynamic phase of the exemplary systems at infinite temperature exhibits multiple discrete values due to the resonant points. In contrast, we show that for a system with a continuous spectrum, no resonant behavior can be found and the dynamic phase does not exhibit multiple discrete values at infinite temperature. There have been studies suggesting quantum behavior at infinite temperature~\cite{Berkelbach10,Nagy17,Roy18,Kemp19}, and our analysis of the dynamic phase of mixed quantum states provides more examples.

The rest of the paper is organized as follows. Sec.~\ref{CB} summarizes the fiber-bundle formalism of the Berry phase and how to construct the combined Berry process for the system to acquire both the Berry phase and the dynamic phase. Sec.~\ref{CU} summarizes the fiber-bundle formalism of the Uhlmann phase and its parallel-transport condition of mixed quantum states. Sec.~\ref{ID} shows the incompatibility between the Uhlmann process and the dynamic process. The dynamic phase of mixed quantum states from the general time-evolution is defined and illustrated by several examples. The lack of classical analogue of the dynamic phase of mixed states is also discussed. Sec.~\ref{Conclusion} concludes our work. The Appendix summarizes the technical details of the fiber-bundle language.

\section{Berry phase in the fiber-bundle language}\label{CB}

\subsection{Berry bundle and parallel-transport condition}
We consider a Hamiltonian $\hat{H}(\mathbf{R})$ depending on a set of parameters which can be collectively written as $\mathbf{R}=(R_1,R_2,\cdots, R_k)$.
Its normalized eigenstates are given by $|n,\mathbf{R\rangle}$. The case with no energy degeneracy is considered here, but the conclusions apply to degenerate cases as well. Without loss of generality, we suppose the system initially stays at the ground state $|0,\mathbf{R}\rangle$.
Assume $\mathbf{R}$ changes continuously as a function of a parameter $t$, so $\mathbf{R}=\mathbf{R}(t)$. Here $t$ may or may not be the time. The instantaneous ground state is expressed as $|0,\mathbf{R}(t)\rangle$. The adiabatic condition of this evolution requires that no level crossing takes place, i.e. the system always stays at the instantaneous ground state. In absence of energy degeneracy, we will simplify $|0,\mathbf{R}\rangle$ as $|\mathbf{R}\rangle$ hereafter.

When the adiabatic process evolves along a closed curve in the parameter space, the system acquires a geometric phase, known as the Berry phase~\cite{Berry84,Bohm03}. The fiber bundle for the Berry phase has been described in, for example, Ref.~\cite{Simon83}, and is summarized in  Appendix~\ref{appBerry} by using the geometry language of Ref.~\cite{Nakahara}.
In brief, a U(1) principle bundle, which is called Berry Bundle here, can be defined as $P(H,\textrm{U(1)})$ where $H$ is the quantum phase space and mathematically is a projective space (see Appendix~\ref{appBerry1}), and U(1) is isomorphic to both the fiber space and the structure group.
The Berry phase is produced when a quantum state is parallel-transported along a loop in the parameter space $M$.
Let $\gamma:[0,1]\rightarrow H$ be such a loop in $H$, satisfying $\gamma(0)=\gamma(1)$. A curve in $P$ given by $\tilde{\gamma}:[0,1]\rightarrow P$ is said to be a horizontal lift of $\gamma$ if $\pi\circ\tilde{\gamma}=\gamma$ where $\pi$ is the projection of the bundle. $\tilde{\gamma}$ may not be a closed loop even if $\gamma$ is.

A connection of a fiber bundle, the Ehresmann connection~\cite{Nakahara} in particular, can be constructed by identifying the horizontal and vertical subspaces. The connection then allows a definition of parallel transport of vectors on the fiber. Appendix~\ref{appBerry2} summarizes the construction of the horizontal and vertical subspaces of the bundle for the Berry phase, and we outline some key points here. Let $\tilde{X}$ be the tangent vector of $\tilde{\gamma}$ and $|\psi(t)\rangle=\me^{\mi\theta(t)}|\mathbf{R}(t)\rangle$ be a point on $\tilde{\gamma}$, then $\tilde{\gamma}$ is a horizontal lift of $\gamma$ if
\begin{align}\label{IHC0}
\langle\psi(t)|\tilde{X}|\psi(t)\rangle=\langle\psi(t)|\frac{\dif_P }{\dif t}|\psi(t)\rangle=0
\end{align}
for any $t$. Here $\dif_P$ is the exterior derivative on $P$.
Moreover,
$\textrm{Re}  \langle \psi(t)|\frac{\dif }{\dif t}|\psi(t)\rangle=0$ due to $\langle \psi(t)|\psi(t)\rangle=1$.
Here the operator $\dif$ is the derivative on $M$ (or equivalently, on H).
Hence the horizontal-space condition for $\tilde{\gamma}$ is
\begin{align}\label{IHC}
\textrm{Im} \langle \psi(t)|\frac{\dif }{\dif t}|\psi(t)\rangle=0.
\end{align}
The condition (\ref{IHC0}) indicates that we can define a $u(1)$-valued one-form, i.e. a connection at $|\psi\rangle$ on $P$ by
\begin{align}\label{CP}
\omega_{|\psi\rangle}=\langle \psi|\dif_P|\psi\rangle.
\end{align}
This is in fact the Ehresmann connection on $P$.
Since $\langle \psi|\dif_P|\psi\rangle$ is imaginary-valued, the expression is equivalent to
$\omega_{|\psi\rangle}=\mi\textrm{Im}\langle \psi|\dif_P|\psi\rangle$. With the introduction of $\omega$, the horizontal-space condition (\ref{IHC0}) can be further expressed as
\begin{align}\label{IHCO}
\omega(\tilde{X})=0.
\end{align}

The Berry connection, which is the pull-back of $\omega$, is defined on the base manifold, or $M$. Let $X$ be the push-forward of $\tilde{X}$, then Eq.~(\ref{CP}) indicates
\begin{align}\label{BcX}
A_{B}(X)=\langle \mathbf{R}(t)|\frac{\dif}{\dif t}|\mathbf{R}(t)\rangle,
\end{align}
By using $|\psi(t)\rangle=\me^{\mi\theta(t)}|\mathbf{R}(t)\rangle$ and Eq.~(\ref{IHCO}), we have
\begin{align}\label{OmegaX}
\langle \mathbf{R}(t)|\frac{\dif}{\dif t}|\mathbf{R}(t)\rangle +\mi\frac{\dif \theta(t)}{\dif t}=0.
\end{align}
and the Berry phase is
\begin{align}\label{Bp}
\theta_B=\theta(1)=\mi\oint A_B(X(t))\dif t.
\end{align}
Here we emphasize that only the closed loops in the parameter space with $R(0)=R(1)$ are considered in the paper.
Since $|\psi(1)\rangle=\me^{\mi\theta_B}|\psi(0)\rangle$, we have
\begin{align}\label{TB}
\theta_B=\arg\langle \psi(0)|\psi(1)\rangle.
\end{align}

The parallel transport on the bundle constructed here can be understood as follows. Eq.~\eqref{OmegaX} leads to Eq.~(\ref{ox}), which can be further expressed as
\begin{align}\label{ox1}
&\nabla_Xg_\gamma(t)\equiv\frac{\dif g_\gamma(t)}{\dif t}+A_{B}(X)g_\gamma(t)=0.
\end{align}
where $\nabla_X$ means taking the covariant derivative along the $X$ direction, i.e. the tangent direction of curve $\gamma(t)$ on $H$. $\nabla_Xg_\gamma(t)=0$ simply reflects that $g_\gamma(t)$, a fiber element, is parallel-transported along $\gamma(t)$. When written explicitly in the components, the parallel transport equation becomes $\frac{\partial g_\gamma(t)}{\partial R^i}+A_{Bi}g_\gamma(t)=0$.
Thus, a parallel transport of a wave-function along $\tilde{\gamma}(t)$ is equivalent to a parallel transport of the associated fiber element $g_\gamma$ along $\gamma(t)=\pi(\tilde{\gamma}(t))$. %Note the parallel-transport condition is traced back to the horizontal-lift condition given by Eq.~(\ref{IHC0}), which is the same as the condition for adiabatic time-evolution. Physically, it means that the change of a quantum state $|\psi(t)\rangle$ when it is parallel-transported is perpendicular to the state itself. Thus, the change does not contain any component of other instantaneous eigenstates of the Hamiltonian, so there is no level crossing during the process. Therefore, the adiabatic condition is equivalent to the geometric, parallel-transport condition.

Furthermore, it can be shown that the parallel-transport condition for the fiber element leads to the concept of ``parallelity" between two quantum pure states as
\begin{align}\label{p1}
\langle\psi_1|\psi_2\rangle=\langle\psi_2|\psi_1\rangle>0.
\end{align}
It is also equivalent to Pancharatnam's notation of parallelity~\cite{Pancharatnam56,Uhlmann86}.
Moreover, parallelity is a symmetric relation but not a transitive one. The failure of transitivity is measured by the Berry curvature.

\subsection{Dynamical Phase and Berry Phase}\label{DPBP}
It is worth noting the parameter $t$ of the loop $\gamma$ may not necessarily be \textit{the time}. If it is chosen as the time, then the corresponding system naturally experiences the time evolution governed by the Schrodinger equation
 \begin{align}\label{Se}
\mi\hbar\frac{\dif}{\dif t}|\psi(t)\rangle=\hat{H}(\mathbf{R}(t))|\psi(t)\rangle,
\end{align}
where $\hat{H}(\mathbf{R}(t))$ is the instantaneous Hamiltonian with the time-dependent parameter $\mathbf{R}$. If the evolution is adiabatic, i.e.,
the changing rate of the parameter is slow enough such that no quantum transitions between different instantaneous states can occur,
the system acquires a dynamic phase $\me^{-\frac{\mi}{\hbar}\int_0^TE_0(t)\dif t}$ after a cycle $T$. The dynamic phase is determined by Eq.~(\ref{Se}), and the Berry phase is the solution to the parallel-transport/horizontal-space condition (\ref{OmegaX}). Otherwise, if $t$ is a parameter different from the time, then $\gamma$ denotes a loop in the parameter space that is not directly related to the dynamics.

If $t$ denotes the time, the evolution of $|n,\mathbf{R}(0)\rangle$ exhibits an interesting property. In the general situation, $|\mathbf{R}(t)\rangle\equiv |n,\mathbf{R}(t)\rangle$ is given by
\begin{align}
 |\mathbf{R}(t)\rangle=\me^{-\frac{\mi}{\hbar}\int_0^tE_n(\tau)\dif \tau}\me^{-\int_0^t\langle n, \mathbf{R}(\tau)|\frac{\partial}{\partial \tau}|n, \mathbf{R}(\tau)\rangle\dif \tau} |\mathbf{R}(0)\rangle\notag.
\end{align}
By this relation and  Eq.~(\ref{psit0}), the  Schrodinger equation (\ref{Se}) gives~\cite{MessiahBook,BrihayePLA94}
\begin{align}\label{Rady}
& \mi\hbar\frac{\dif }{\dif t}|n, \mathbf{R}(t)\rangle=\mi\hbar\frac{\partial}{\partial t}|n, \mathbf{R}(t)\rangle+ \notag\\
 &\big[E_n-\mi\hbar\langle n, \mathbf{R}(t)|\frac{\partial}{\partial t}|n, \mathbf{R}(t)\rangle\big]|n, \mathbf{R}(t)\rangle\notag\\
 &=(\hat{H}+\mi\hbar\hat{K})|n, \mathbf{R}(t)\rangle.
\end{align}
Here $\hat{K}=\sum_m \dot{\hat{P}}_m(t) \hat{P}_m(t)$ with $\hat{P}_m(t)=|m,\mathbf{R}(t)\rangle\langle m,\mathbf{R}(t)|$ being the projector onto the state $|m,\mathbf{R}(t)\rangle$. Note that $\hat{K}^\dagger=-\hat{K}$ due to $\sum_n \hat{P}_n=\sum_n\hat{P}^2_n=1$.
The result can be generalized to quantum systems with degenerate energy levels. Assuming the degeneracy of the $n$-th level is $N_n$ and the corresponding state is $|n\rangle_a$ with $a=1,\cdots,N_n$, Eq.~(\ref{Rady}) becomes
$ \mi\hbar\frac{\dif }{\dif t}|n, \mathbf{R}(t)\rangle_a=\sum_b(\hat{H}+\mi\hbar\hat{K})_{ab}|n, \mathbf{R}(t)\rangle_b$,
where $\hat{K}=\sum_{m,b} \dot{\hat{P}}_{m,b}(t) \hat{P}_{m,b}(t)$ with $\hat{P}_{m,b}(t)=|m,\mathbf{R}(t)\rangle_b {}_b\langle m,\mathbf{R}(t)|$.

One may introduce the ``adiabatic Hamiltonian" $\hat{H}_\text{ad}\equiv\hat{H}+\mi\hbar\hat{K}$ to describe the adiabatic dynamics of any time-dependent observable in the Heisenberg picture~\cite{BrihayePLA94}.
Importantly, the adiabatic evolution of the corresponding density matrix $\rho_n=|n, \mathbf{R}(t)\rangle\langle n, \mathbf{R}(t)|$ follows
 \begin{align}\label{deofr}
\dot{\rho_n}=-\frac{\mi}{\hbar}[\hat{H}_\text{ad},\rho_n].
\end{align}
We call the process governed by $\hat{H}_\text{ad}$ the adiabatic dynamic process, or simply the Berry process because the Berry phase is produced during this process.

Since the density matrix can describe mixed quantum states as well, a fundamental question is whether there exists any generalization of the adiabatic dynamic equation to mixed quantum states. Moreover, the dynamical phase is accumulated by a pure quantum state evolving with time according to the Schrodinger equation. Another question is then what is the dynamical phase produced by the time evolution of mixed quantum states? Investigating those questions may help us understand the physical implications of the time evolution of mixed quantum states.

\section{Mixed quantum states and Uhlmann phase in fiber-bundle language}\label{CU}
\subsection{Mixed quantum states, Uhlmann bundle, and Uhlman phase}
To generalize the previous discussions to mixed quantum states, we notice that the base manifold of the Berry bundle is a projective Hilbert space formed by the rank-one density matrices $\rho=|\psi\rangle\langle \psi|$. A similar construction of the Uhlmann bundle and Uhlmann holonomy for mixed quantum states can be performed.
As shown in Appendix~\ref{appBerry1}, the $U(1)$ transformation leads to an equivalent relation between the pure quantum states in a Hilbert space, causing a redundancy in the determination of a pure physical state. Similar discussions can be constructed for mixed quantum states. Following Uhlmann's approach,
an operator $W$ is called the ``amplitude" of a density matrix $\rho$ and a unitary matrix $U$ is called a ``phase factor of $\rho$" if and only if
\begin{align}
\rho=WW^\dagger,\qquad W=\sqrt{\rho}U.
\end{align}
The phase factor comes from the unique polar decomposition of amplitude if the density matrix is full-ranked.
The key idea of Uhlmann's approach is to lift the action of $\rho$ to an extended Hilbert space. This can be most clearly illustrated by borrowing the terminology from quantum information known as the purification of the density matrix. $W$ is said to be a purification of $\rho$, or $W$ purifies $\rho$. The details are summarized in Appendix~\ref{appUhlmann1}.

The geometrical description of the Berry phase can then be generalized to the Uhlmann phase.
The Uhlmann bundle~\cite{ViyuelaPRL14-2} is constructed as $(E,\pi,Q,F,\textrm{U}(n))$. Here $E$ is the total space, and $\pi$ is the projection acting as $\pi: E\rightarrow Q$
 \begin{align}
 \pi(W)=WW^\dagger=\rho.
\end{align}
$Q$ is the base space formed by the full rank density matrix $\rho$. We assume $M$ is a manifold of the parameter $\mathbf{R}$ that parametrizes $Q$. There are some subtleties. For example, a linear combination of the density matrices may not produce a valid density matrix. However, those issues may be circumvented with suitable constraints~\cite{Asorey19}. $F$ is the fiber, i.e. the Hilbert space spanned by the amplitudes, which can be denoted either by $H_W$ or by $\mathcal{H}\otimes\mathcal{H}$. U$(n)$ is the structure group, of which the element acts on the fiber. It can be shown~\cite{Asorey19} that $F$ is diffeomorphic to U$(n)$, so the Uhlmann bundle is considered as a principle bundle. It was found~\cite{DiehlPRB15} that the Uhlmann bundle is a trivial bundle since it admits a global section $\sigma(\rho)=\sqrt{\rho}$.

The separation of the tangent bundle $TE$ can be achieved by introducing a connection $\omega$ on $E$, which projects $TE$ onto $HE$. Specifically, if $\tilde{X}$ is a horizontal vector, then\begin{align} \label{OmegaXU}\omega(\tilde{X})=0,\end{align}
which is the generalization to Eq.~(\ref{OmegaX}) or Eq.(\ref{CPHC}). Here we emphasize again that the connection one-form $\omega$ is the Ehresmann connection~\cite{Nakahara}, i.e., it separates $TE_W$ into $HE_W\oplus VE_W$ via Eq.~(\ref{OmegaXU}). Following Ref.~\cite{Uhlmann91,GPbook}, one can introduce $\omega$ via
\begin{align}\label{UC3}
W^\dagger\dif W-\dif W^\dagger W=W^\dagger W\omega+\omega W^\dagger W.
\end{align}
It can be shown that under the gauge transformation $W'=WV$, the connection transforms properly as $\omega'=V^\dagger \omega V+V^\dagger\dif V$~\cite{GPbook}.

The Uhlmann connection is the pull-back of $\omega$ and can be derived from Eq.~(\ref{UC3}). It takes the form
\begin{align}\label{AU9}
A_U=-\dif UU^\dagger,
\end{align}
where $\dif$ should be understood as the ``horizontal lift'' of the exterior derivative on $Q$, i.e., it does not contain the component of the derivative in the fiber space. The detailed derivation is given in Appendix~\ref{appUhlmann3}. This expression of the Uhlmann connection agrees with that of Ref.~\cite{DiehlPRB15}, but both differ from that of Ref.~\cite{ViyuelaPRL14} by a minus sign. The difference in the sign may lead to different results because the connection affects the evolution of the density matrix during the Uhlmann process, which will be discussed later.

The explicit expression of the Uhlmann connection is
\begin{align}\label{AUE}
A_U=\sum_{ij}|i\rangle\langle i|A_U|j\rangle\langle j|=-\sum_{ij}|i\rangle\frac{\langle i|[\mathrm{d}\sqrt{\rho},\sqrt{\rho}]|j\rangle}{\lambda_i+\lambda_j}\langle j|,
\end{align}
where $\lambda_i$ and $|i\rangle$ are the eigenvalue and eigenvector of the density matrix.
The derivation is given in Appendix~\ref{appUhlmann3}.
Eq.~(\ref{AU9}) indicates that $A_U$ is a pure gauge, so the Uhlmann curvature $F_U=\dif A_U+A_U\wedge A_U$ vanishes. This is consistent with the fact that the Uhlmann bundle is a trivial bundle.
Similar to Eq.~(\ref{TB}) in the pure state case, the Uhlmann phase is given by
\begin{align}\label{Up1}
\theta_U:&=\arg\langle W(0)|W(1)\rangle=\arg\textrm{Tr}[W(0)^\dagger W(1)] \notag \\
&=\arg\textrm{Tr}[\rho(0)\mathcal{P}\me^{-\oint A_U}],%\notag\\
\end{align}
where $W(0)$ and $W(1)$ are the initial and final amplitudes, respectively, and Eq.~(\ref{W8}) has been applied.

Similarly, Eq.~(\ref{AU9}) can be rewritten, after the manipulation show in Appendix~\ref{appUhlmann4}, as
\begin{align}\label{d01}
\nabla_XU\equiv\frac{\dif U}{\dif t}+A_U(X)U=0.
\end{align}
Here $\nabla_X$ is the covariant derivative along the $X$ direction. Eq.~(\ref{d01}) means the phase factor $U(t=1)$ is obtained by a parallel transport of $U(t=0)$ along $\tilde{\gamma}$. For the amplitudes, the corresponding parallel-transport condition is given by
\begin{align}\label{d08}
\dot{W}^\dagger W=W^\dagger \dot{W}.
\end{align}
This can be inferred from Eq.~(\ref{upcd4}), and we omit the subscript ``H" here. By integrating both sides along a curve $\tilde{\gamma}$ with respect to Eq.~(\ref{d01}), we get
\begin{align}\label{upcd6}
W^\dagger(1) W(0)=W(0)^\dagger W(1)>0.
\end{align}
Here ``$>0$'' means that all of the eigenvalues of the matrix are positive, which is because all the eigenvalues of $\rho(0)$ and $\rho(1)$ are positive when they are both full-ranked. This is a generalization of Eq.~(\ref{p1}), which is the parallel transport condition for pure quantum states.

\subsection{Dynamical Process and Uhlmann Process}
Geometrically, the meaning of the parallel-transport condition (\ref{upcd6})
is unequivocal. However, its physical nature is unclear. Previously, we have shown that a pure state evolves according to Eq.~(\ref{Rady}) during an adiabatic dynamic process. In an Uhlmann process, the amplitude, though not uniquely determined, plays a similar role as the quantum pure state. This has been pointed out in the discussion of quantum purification (see Appendix~\ref{appUhlmann1}).
Interestingly, the parallel transport condition for the amplitude can be cast into the form of a differential equation that the amplitude follows. It can be formally written as
\begin{align}\label{Ha}
\mi\hbar\dot{W}=\tilde{H}W
\end{align}
by introducing the an auxiliary matrix $\tilde{H}=\mi\hbar\dot{W}W^{-1}$.
This can be verified by
\begin{align}\label{Ha3}
\dot{W}^\dagger W=\frac{\mi}{\hbar}W^\dagger \tilde{H}^\dagger W=-\frac{\mi}{\hbar}W^\dagger \tilde{H} W=W^\dagger \dot{W}.
\end{align}

If the parameter $t$ is chosen as the time, Eq.~(\ref{Ha}) looks like an ``anti-Hermitian Schrodinger equation", which formally describes the dynamics of the purification $W$. However, Eq.~(\ref{d08}) indicates that $\tilde{H}$ is anti-Hermitian.
Since $\rho=WW^\dagger$, it can be shown that
\begin{align}\label{Urhod}
\dot{\rho}=\dot{W}W^\dagger+W\dot{W}^\dagger=-\frac{\mi}{\hbar}\{\tilde{H},\rho\}.
\end{align}
This defines the dynamics of the density matrix during an Uhlmann process, equivalent to a parallel transport, under which the phase factor changes according to
$\dot{U}U^{-1}=A_U(X)$.
Substituting the result into Eq.~(\ref{Ha}), we have
\begin{align}\label{Ht1}
-\frac{\mi}{\hbar}\tilde{H}\sqrt{\rho}U=\dot{W}
%=\dot{\sqrt{\rho}}U+\sqrt{\rho}\dot{U}
=\dot{\sqrt{\rho}}U-\sqrt{\rho}A_U(X)U.
\end{align}
Therefore,
\begin{align}\label{ahH}
\tilde{H}=\mi\hbar\big[\dot{\sqrt{\rho}}\sqrt{\rho^{-1}}-\sqrt{\rho}A_U(X)\sqrt{\rho^{-1}}\big].
\end{align}

The Uhlmann phase can also be expressed by $\tilde{H}$ since $A_U(X)$ can be obtained from the solution to Eq.~(\ref{ahH}). Explicitly,
\begin{align}
A_U(X)&=\frac{\mi}{\hbar}\sqrt{\rho^{-1}}\tilde{H}\sqrt{\rho}+\sqrt{\rho^{-1}}\dot{\sqrt{\rho}},\label{srhod2a} \\
U(1)&=\mathcal{P}\me^{-\int_0^1\big(\frac{\mi}{\hbar}\sqrt{\rho^{-1}}\tilde{H}\sqrt{\rho}+\sqrt{\rho^{-1}}\dot{\sqrt{\rho}}\big)\dif t}U(0).\label{srhod2}
\end{align}
Since $A_U(X)$ changes under gauge transformations, $\tilde{H}$ also has a gauge degree of freedom. This extra gauge redundancy can be removed only when the Uhlmann process is a closed cycle, i.e., when $\mathbf{R}(1)=\mathbf{R}(0)$. Under the condition, the term $ \frac{\mi}{\hbar}\oint\sqrt{\rho^{-1}}\tilde{H}\sqrt{\rho}\dif t$ appearing in Eq.~(\ref{srhod2}) is gauge independent.

On the other hand, if the parameter $t$ is chosen as the time, the density matrix follows the evolution governed by the Hamiltonian $\hat{H}$ according to~\cite{Merzbacher_book}
\begin{align}\label{rhoe}
\dot{\rho}=-\frac{\mi}{\hbar}[\hat{H},\rho].
\end{align}
Similar to the previous discussion, this equation can also be realized by imposing the ``Schrodinger equation'' for the amplitudes by
\begin{align}\label{Ha1}
\mi\hbar\dot{W}=\hat{H}W.
\end{align}
This defines the dynamic evolution of a mixed quantum mixed. During the dynamic process, a phase factor different from the Uhlmann phase is accumulated. We call it the dynamic phase of a mixed quantum state.

The expression of the dynamics phase can be obtained as follows. Similar to Eq.~(\ref{Ht1}), we get
\begin{align}\label{Ht2}
-\frac{\mi}{\hbar}\hat{H}\sqrt{\rho}U=\dot{W}=\dot{\sqrt{\rho}}U+\sqrt{\rho}\dot{U}
\end{align}
by substituting $W=\sqrt{\rho}U$ into Eq.~(\ref{Ha1}).
This implies that
\begin{align}\label{Ht3}
\dot{U}U^{-1}=-\frac{\mi}{\hbar}\sqrt{\rho^{-1}}\hat{H}\sqrt{\rho}-\sqrt{\rho^{-1}}\dot{\sqrt{\rho}}.
\end{align}
After integrating both sides, we have
\begin{align}\label{UPd}
U(1)=\mathcal{T}\me^{-\int_0^1\big(\frac{\mi}{\hbar}\sqrt{\rho^{-1}}\hat{H}\sqrt{\rho}+\sqrt{\rho^{-1}}\dot{\sqrt{\rho}}\big)\dif t}U(0),
\end{align}
where $\mathcal{T}$ is the time ordering operator. The expression is quite similar to Eq.~(\ref{srhod2}) of the accumulated phase during an Uhlmann process.
It can be thought of as the dynamical phase obtained by the amplitude $W$ during a cyclic time evolution governed by the Hamiltonian.

\section{Incompatibility and Dynamic phase of mixed quantum states}\label{ID}
\subsection{Incompatibility between the two processes}
Now we come back to the question on whether there exists a generalization of the Berry process given by Eq.~(\ref{deofr}) to mixed quantum states. Unfortunately, Eq.~(\ref{Urhod}) corresponding to the Uhlmann process is not compatible with the dynamical equation (\ref{rhoe}). If one combines them, it will have a structure known to violate conservation of probability, causing $Tr(\rho)=1$ to fail, and lose invariance against a shift of the zero-energy~\cite{Serg14}. This is because
the right-hand-side of the former is an anti-commutator, but that of the latter is a commutator. This is in start contrast to Eq.~(\ref{deofr}) of the time-dependent Berry process.

One may argue that Eqs.~(\ref{Urhod}) and (\ref{rhoe}) can be made compatible if one necessary condition is established: The right-hand-side of ether one of them vanishes. Here we analyze the two scenarios and show that none of them is valid.

\subsubsection{$\{\tilde{H},\rho\}=0$}
The condition $\{\tilde{H},\rho\}=0$ is not possible. Otherwise, the combination of $\tilde{H}$ being anti-Hermitian and $\tilde{H}\rho=-\rho\tilde{H}$ implies
\begin{align}\label{UPd7}
\sqrt{\rho^{-1}}\tilde{H}\sqrt{\rho}=-\sqrt{\rho}\tilde{H}\sqrt{\rho^{-1}}=\left(\sqrt{\rho^{-1}}\tilde{H}\sqrt{\rho}\right)^\dagger,
\end{align}
i.e., $\sqrt{\rho^{-1}}\tilde{H}\sqrt{\rho}$ is Hermitian. If $\rho$ is full-ranked, $W=\sqrt{\rho}U$ and
$\tilde{H}=\mi\hbar\dot{W}W^{-1}$ are full-ranked. Since $\tilde{H}$ is anti-Hermitian, all of its eigenvalues must be purely imaginary. Moreover, $\sqrt{\rho^{-1}}\tilde{H}\sqrt{\rho}$ is a similarity transformation of $\tilde{H}$, hence it must have the same eigenvalues as the latter. However, Eq.~(\ref{UPd7}) implies that all eigenvalues of $\sqrt{\rho^{-1}}\tilde{H}\sqrt{\rho}$ are real. Thus, we reach a contradiction.

\subsubsection{$[\hat{H},\rho]=0$}
Next, we consider the situation with $[\hat{H},\rho]=0$, i.e., $\dot{\rho}=0$, which includes the quasistatic processes. If the parameter $t$ is the time, a necessary condition for realizing the Uhlmann process is to keep the system at equilibrium at any time, consistent with the quasistatic condition.
%Hence, the mixed quantum state does not need to undergo an adiabatic process.
In this sense, the Uhlmann phase is more like a generalization of the Aharonov-Anandan phase~\cite{AAphase}.
In fact, the examples of the Ulhmann phase given in Ref.~\cite{ViyuelaPRL14} all belong to this situation since the equilibrium density matrix $\rho=\frac{1}{Z}\me^{-\beta\hat{H}}$, where $\beta=\frac{1}{k_BT}$ with $k_B$ being the Boltzmann constant and $Z$ the partition function, was used in the derivations.

However, we need to carefully check if the condition $[\hat{H},\rho]=0$ can be imposed without causing problems. Since $\rho=WW^\dagger$, the condition implies that
\begin{align}\label{UPd8}
\dot{W}W^\dagger+W\dot{W}^\dagger=0.
\end{align}
If the expression is compatible with the Uhlmann parallel-transport condition given by Eq.~(\ref{d08}), or equivalently Eq.~(\ref{Ha}), the left-hand-side of Eq.~(\ref{UPd8}) will become
\begin{align}\label{UPd9}
0=-\frac{\mi}{\hbar}\tilde{H}WW^\dagger+\frac{\mi}{\hbar}WW^\dagger\tilde{H}^\dagger=-\frac{\mi}{\hbar}\{\tilde{H},\rho\},
\end{align}
where the anti-Hermitian property $\tilde{H}^\dagger=-\tilde{H}$ has been applied. This proves that the equilibrium condition $[\hat{H},\rho]=0$ cannot coexist with the condition $\{\tilde{H},\rho\}\neq 0$.

Hence, choosing the parameter $t$ as the time again leads to a contradiction, so the Uhlmann process is not compatible with the dynamic process governed by the Hamiltonian. In other words, a mixed quantum state cannot obtain the Uhlmann phase and dynamic phase simultaneously during a single process. Even more, the ``process" does not necessarily need to be parameterized by the time. For an Uhlmann process, the system acquires the Uhlmann phase given by Eq.~(\ref{Up1}), but for a dynamic process, the system acquires the dynamic phase given by
\begin{align}\label{Ud1}
\theta_D:&=\arg\langle W(0)|W(1)\rangle=\arg\textrm{Tr}[W(0)^\dagger W(1)] \notag \\
&=\arg\text{Tr}[\rho(0)\mathcal{T}\me^{- \mathlarger{\oint}\big(\frac{\mi}{\hbar}\sqrt{\rho^{-1}}\hat{H}\sqrt{\rho}+\sqrt{\rho^{-1}}\dot{\sqrt{\rho}}\big)\dif t}].
\end{align}
The phase thus depends on the underlying process.

\subsection{Dynamic phase from quasi-static dynamic process}
Now we focus on the properties of the dynamic phase of mixed quantum states. If $t$ denotes the time, $\mi\hbar\dot{\rho}=[\hat{H},\rho]\neq 0$ corresponds to a non-equilibrium process. The associated dynamic phase is evaluated by the most general formula (\ref{Ud1}). For simplicity, here we consider the class of quasi-static processes with $[\hat{H},\rho]=0$. Moreover, an arbitrary dynamic process may not be periodic with $\rho(0)=\rho(1)$, but here we focus on the cyclic process to study the difference between the Uhlmann phase and the dynamic phase. The equilibrium condition $[\hat{H},\rho]=0$ implies $[\hat{H},\sqrt{\rho}]=0$, we then get
 \begin{align}\label{Ud2}
\theta_D=\arg\text{Tr}[\rho(0)\mathcal{T}\me^{-\frac{\mi}{\hbar} \mathlarger{\oint}\hat{H}\dif t- \mathlarger{\oint}\sqrt{\rho^{-1}}\dot{\sqrt{\rho}}\dif t}].
\end{align}
The second exponent vanishes according to
\begin{align}\label{Ud3}
\oint\sqrt{\rho^{-1}}\dif\sqrt{\rho}=-\ln\sqrt{\frac{\rho(0)}{\rho(0)}}=0.
\end{align}
Thus, we finally get
\begin{align}\label{Ud4}
\theta_D=\arg\text{Tr}[\rho(0)\mathcal{T}\me^{-\frac{\mi}{\hbar} \mathlarger{\oint}\hat{H}\dif t}].
\end{align}
The dynamic phase of a mixed quantum state is the generalization of that of a pure quantum state.

One may argue that if $[\hat{H},\rho]=0$, $\dot{\rho}=0$. Then, $\rho$ does not change with time and there is no accumulated dynamic phase. However, for a quasi-static process, $\dot{\rho}$ is not set to exactly zero for every instance of time. In a realistic situation, only $\dot{\rho}\approx0$ can be held. Thus, $\rho$ gradually changes with time, and the system acquires a dynamic phase over a long time. Similar arguments have been used in thermodynamics textbooks~\cite{Callen_book,Schroeder_book}.

If the Hamiltonian is independent of time, one further obtains
\begin{align}\label{Ud41}
\theta_D=\arg\left(\sum_n\langle n|\rho(0)|n\rangle\me^{-\frac{\mi}{\hbar}E_n\tau}\right).
\end{align}
Here $\tau$ (instead of $1$ hereafter to emphasize the unit) is the time duration of the cyclic process, and the trace is taken over the Hilbert space spanned by the eigenvectors $\{|n\rangle\}$ of the Hamiltonian.
The expression indicates that the dynamic phase of a mixed quantum state is not simply the weighted sum of the dynamic phases of its constituent states. The latter is given by \begin{align}\label{t1}-\sum_np_n\frac{E_n\tau}{\hbar}\quad \text{with}\quad p_n=\langle n|\rho(0)|n\rangle.\end{align}
Eq.~(\ref{Ud41}) indicates that the interference effect between the different constituent states survives even though the system is a mixed quantum state.
This is different from the expectation value of an observable $\hat{O}$ in a mixed quantum state, given by
\begin{align}\label{t2}
\langle\hat{O}\rangle=\sum_np_n\langle n|\hat{O}|n\rangle.
\end{align}
One can see that Eq.~(\ref{t1}) is of such a structure, but the dynamic phase \eqref{Ud41} is not. We caution that the phase is not an observable in quantum mechanics although it may cause interference of the wavefunctions.

For the ground state $|n_0\rangle$, $\rho(0)=|n_0\rangle\langle n_0|$.  Eq.~(\ref{Ud41}) reduces to the known results from quantum mechanics~\cite{Sakurai_QM}:
\begin{align}\label{Ud5}
\theta_D=\arg(\me^{-\frac{\mi}{\hbar}E_{n_0}\tau})=-\frac{E_{n_0}\tau}{\hbar} \mod 2\pi.
\end{align}
For general situations, including non-equilibrium processes, the dynamical phase of mixed quantum systems must be evaluated according to Eq.~(\ref{UPd}). We remark that although the dynamic phase does not reveal the underlying topological information, the definition and calculation will help future research on topological properties of mixed quantum states because one will know how to subtract the non-topological contribution from the dynamic phase.

\subsection{Examples}
Unlike the Uhlmann phase, the dynamic phase does not carry topological information since it depends on the evolution path. In the following, we present two examples of one-dimensional (1D) two-band fermionic systems with periodic boundary condition and compare the result to the harmonic oscillator. We mention that the Uhlmann phase of several 1D two-band models has been studied in Ref.~\cite{ViyuelaPRL14}, and we do not repeat the results here.

\subsubsection{1D two-band models}
We consider 1D Hamiltonians of the quadratic form $\hat{H}=\sum_k\Psi_k^\dagger H_k\Psi_k$, where $\Psi_k=(a_k,b_k)^T$ stands for the two-component fermion operators. $\mathbf{k}\equiv k$ is the crystal momentum in the first Brillouin zone. Moreover,
\begin{align}\label{Hk}
H_k=f(k)1_{2\times2}+\frac{1}{2}\Delta_k\vec{\sigma}\cdot\hat{\mathbf{n}}_k.
\end{align}
Here $\vec{\sigma}=(\sigma_x,\sigma_y,\sigma_z)^T$ denotes the Pauli matrices, and $\hat{\mathbf{n}}_k=(\sin\theta_k\cos\phi_k,\sin\theta_k\sin\phi_k,\cos\theta_k)^T$.
The density matrix in the canonical ensemble is given by
\begin{align}\label{dm}
\rho_k=\frac{\me^{-\beta H_k}}{\text{Tr}(\me^{-\beta H_k})}=\frac{1}{2}\left(1-\tanh\frac{\beta\Delta_k}{2}\vec{\sigma}\cdot\hat{\mathbf{n}}_k\right),
\end{align}
where $\beta=\frac{1}{k_B T}$ with $k_B$ being the Boltzmann constant.
A simple choice of the parametrization of a loop in momentum space is $k(t)=2\pi \frac{t}{\tau}$. The period $\tau$ depends on the model and will be explicitly given later on. As the system evolves with time, $k$ goes through the whole Brillouin zone, which has the shape of a circle $S^1$, if $t$ runs from 0 to $\tau$. For a quasi-static process, the dynamic phase is then given by
 \begin{align}\label{Ud6}
\theta_D&=\arg\text{Tr}[\rho(0)\mathcal{T}\me^{-\frac{\mi}{\hbar} \mathlarger{\oint}H_{k(t)}\dif t}]\notag\\&
%=\arg\text{Tr}[\rho(0)\me^{-\frac{\mi}{2\pi\hbar} \mathlarger{\int}_0^{2\pi}H_{k}\tau\dif k}]\notag\\&
=\arg\text{Tr}[\rho(0)\me^{-\frac{\mi}{2\pi\hbar}\mathlarger{\int}_0^{2\pi}f(k)\tau\dif k}\me^{-\frac{\mi}{2\pi\hbar}\mathlarger{\int}_0^{2\pi}\frac{1}{2}\Delta_k\vec{\sigma}\cdot\hat{\mathbf{n}}_k\tau\dif k}].
\end{align}
One can check the above expression gives the ground-state dynamic phase as $T\rightarrow 0$.

We first consider the periodic Kitaev chain~\cite{Kitaev} with the Hamiltonian
 \begin{align}\label{KH}
\hat{H}=\sum_{i=1}^L(-Ja^\dagger_ia_{i+1}+Ma_ia_{i+1}-\frac{\mu}{2}a^\dagger a_{i}+\text{H.c.}).
\end{align}
Here $L$ is the number of sites, $J$ is the hopping coefficient, $\mu$ is the chemical potential, and $M>0$ is the superconducting gap. We introduce $m=\frac{\mu}{2M}$ and $c=\frac{J}{M}$ and use the Nambu spinor $\Psi_k=(a_k,a_{-k}^\dagger)^T$. The Hamiltonian can be expressed in the form of Eq.~(\ref{Hk}) in momentum space with $\Delta_k=2M\sqrt{(c\cos k-m)^2+\sin^2k}$ and $\hat{\mathbf{n}}_k=\frac{2M}{\Delta_k}(0,-\sin k,-m+c\cos k)^T$~\cite{ViyuelaPRL14}. By Eq.~(\ref{Ud6}), we get
\begin{align}\label{Ud7}
\theta_D&=\arg\text{Tr}[\rho(0)\me^{\mi m\sigma_z\frac{M\tau}{\hbar}}]\notag\\
&=\arg\left[\cos (\frac{mM\tau}{\hbar})+\mi\sin( \frac{mM\tau}{\hbar})\text{Tr}\left(\rho(0)\sigma_z\right)\right],
\end{align}
where $\rho(0)\equiv\rho_{k(0)}$. Note that $\Delta_{k(0)}=2M|c-m|$ and $\hat{\mathbf{n}}_{k(0)}=(0,0,\text{sgn}(c-m))^T$ with $\text{sgn}(c-m)=\frac{c-m}{|c-m|}$. By using Eq.~(\ref{dm}), we have
\begin{align}\label{dm1}
\rho(0)=\frac{1}{2}\left[1-\tanh(\beta M(c-m))\sigma_z\right].
\end{align}
Here we have applied  $\tanh(\beta M|c-m|)\text{sgn}(c-m)=\tanh(\beta M(c-m))$. After substituting Eq.~(\ref{dm1}) into Eq.~(\ref{Ud7}), we finally get
\begin{widetext}
\begin{align}\label{Ud8}
\theta_D&=\arg\left[\cos (\frac{mM\tau}{\hbar})-\mi\sin( \frac{mM\tau}{\hbar})\tanh(\beta M(c-m))\right]\notag\\
&=\left\{\begin{array}{cc}
-\arctan\left[\tan(\frac{mM\tau}{\hbar})\tanh(\beta M(c-m))\right], & \text{if  } \frac{mM\tau}{\hbar} \in \left(2n\pi-\frac{\pi}{2},2n\pi+\frac{\pi}{2}\right);\\
\mp\frac{\pi}{2}, & \text{if  } \frac{mM\tau}{\hbar}=2n\pi\pm\frac{\pi}{2};\\
-\arctan\left[\tan(\frac{mM\tau}{\hbar})\tanh(\beta M(c-m))\right]-\pi, & \text{if  } \frac{mM\tau}{\hbar} \in \left(2n\pi+\frac{\pi}{2},2n\pi+\pi\right);\\
-\arctan\left[\tan(\frac{mM\tau}{\hbar})\tanh(\beta M(c-m))\right]+\pi, & \text{if  } \frac{mM\tau}{\hbar} \in \big[2n\pi-\pi,2n\pi-\frac{\pi}{2}\big),
\end{array}\right.
\end{align}
\end{widetext}
where $n$ is an arbitrary integer, and the range of the dynamic phase is $(-\pi,\pi]$.

Next, we consider the periodic Su-Schrieffer-Heeger (SSH) model~\cite{SSH}, which is a lattice with alternating hopping coefficients described by the Hamiltonian
\begin{align}\label{SSH}
\hat{H}=\sum_{i=1}^L(J_1a_i^\dagger b_i+J_2a_i^\dagger b_{i-1}+\text{H.c.})+V\sum_{i=1}^L(a^\dagger_ia_i-b^\dagger_ib_i).
\end{align}
Here we consider the situation with $V=0$ and $J_2>J_1$. When expressed in momentum space, the Hamiltonian is in the form of Eq.~(\ref{Hk}) with
$\Delta_k=2 \sqrt{J^2_1+J^2_2+2J_1J_2\cos k}$ and $\hat{\mathbf{n}}_k=\frac{2 }{\Delta_k}(-J_1-J_2\cos k,J_2\sin k,0)^T$. Following the same steps, it can be found that $\Delta_{k(0)}=2|J_1+J_2|$, $\hat{\mathbf{n}}_{k(0)}=(-\text{sgn}(J_1+J_2),0,0)^T$ and
\begin{align}\label{dm2}
\rho(0)=\frac{1}{2}\left[1+\tanh(\beta(J_1+J_2))\sigma_x\right].
\end{align}
The dynamic phase is given by
\begin{widetext}
\begin{align}\label{Ud9}
\theta_D&=\arg\text{Tr}[\rho(0)\me^{\mi \frac{J_1\tau}{\hbar}\sigma_x}]\notag\\
%&=\arctan\left[\tan (\frac{J_1\tau}{\hbar})\text{Tr}\left(\rho(0)\sigma_x\right)\right]\notag\\
&=\left\{\begin{array}{cc}\arctan\left[\tan (\frac{J_1\tau}{\hbar})\tanh(\beta(J_1+J_2))\right], & \text{if } \frac{J_1\tau}{\hbar} \in  \left(2n\pi-\frac{\pi}{2},2n\pi+\frac{\pi}{2}\right);\\
\pm\frac{\pi}{2}, & \text{if  } \frac{J_1\tau}{\hbar}=2n\pi\pm\frac{\pi}{2};\\
\arctan\left[\tan (\frac{J_1\tau}{\hbar})\tanh(\beta(J_1+J_2))\right]+\pi, & \text{if  } \frac{J_1\tau}{\hbar} \in \left(2n\pi+\frac{\pi}{2},2n\pi+\pi\right);\\
\arctan\left[\tan (\frac{J_1\tau}{\hbar})\tanh(\beta(J_1+J_2))\right]-\pi, & \text{if  } \frac{J_1\tau}{\hbar} \in \big[2n\pi-\pi,2n\pi-\frac{\pi}{2}\big).
\end{array}\right.
\end{align}
\end{widetext}

\begin{figure}[th]
\centering
\includegraphics[width=2.5in,clip]
{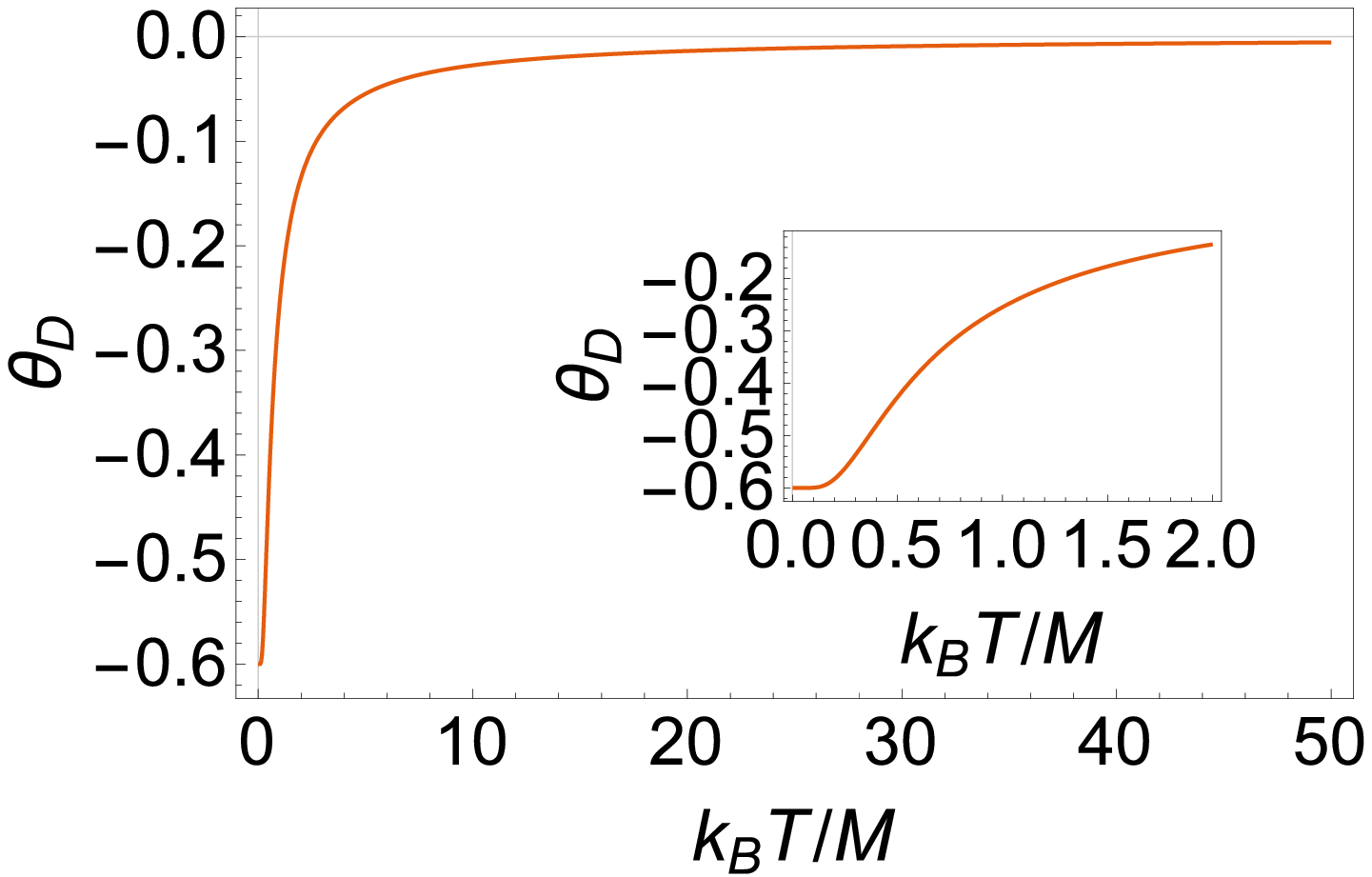}
\includegraphics[width=2.5in,clip]
{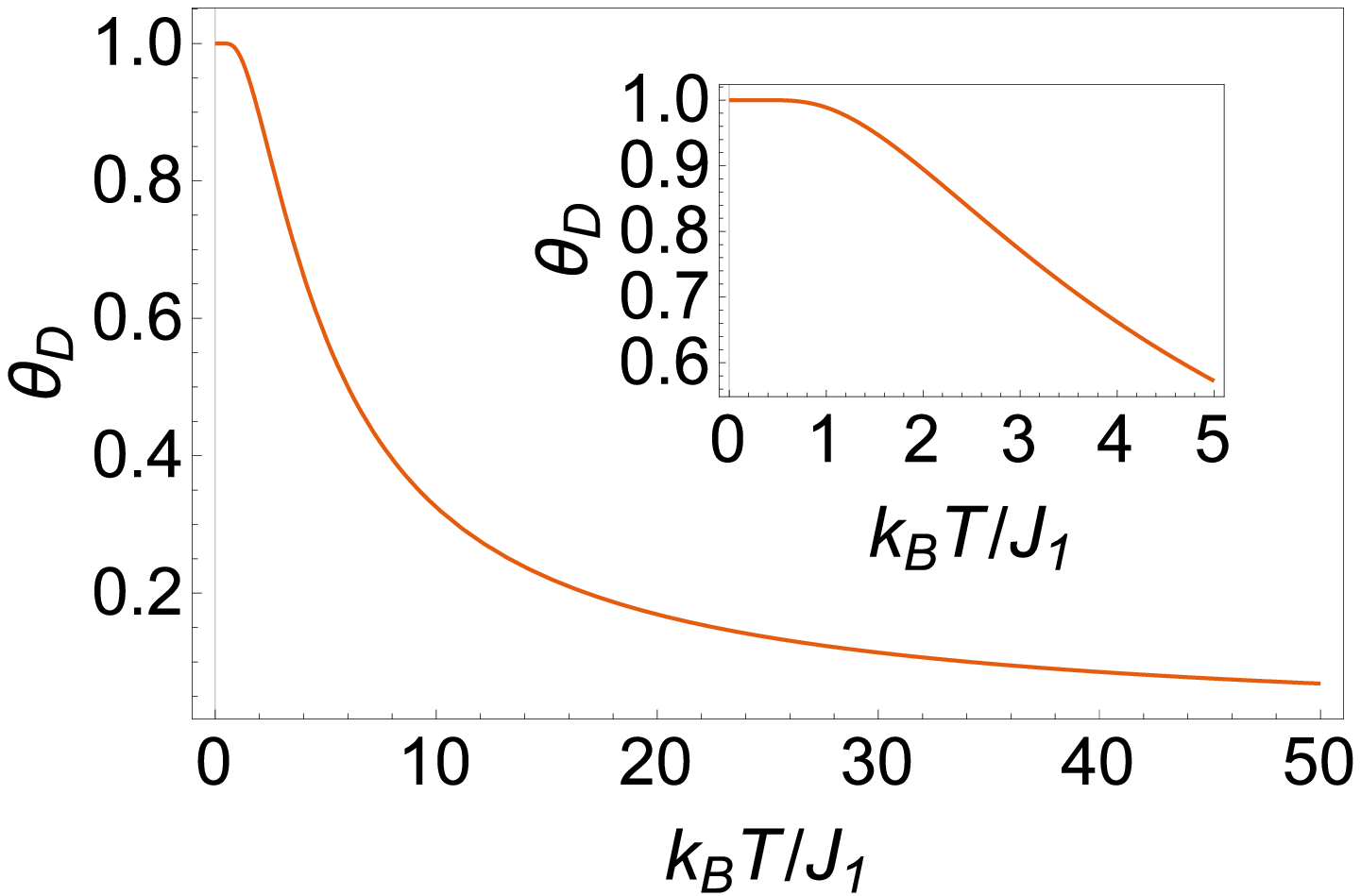}
\caption{Dynamic phase as a function of temperature. The top (bottom) panel shows the result of the Kitaev chain (SSH model). The insets show the detailed structures when the temperature is relatively low. The parameters for the Kitaev chain (SSH model) are $m=0.6$ and $c=1.0$ ($J_2/J_1=1.2$). The time duration $\tau$ is chosen such that $\frac{M\tau}{\hbar}=1$ for the Kitaev chain and $\frac{J_1\tau}{\hbar}=1$ for the SSH model.}
\label{Fig1}
\end{figure}

Figure.~\ref{Fig1} shows the dynamic phase of the two models as a function of temperature. We set $\frac{M\tau}{\hbar}=\frac{J_1\tau}{\hbar}=1.0$ to fix the units of time.
For the Kitaev chain, $c=1$ and $m=0.6$. For the SSH model, $J_2=1.2$ in units of $J_1$. Clearly, the dynamic phase is not quantized for both situations, and only continuous curves are observed. In our calculations, periodic boundary condition is used to utilize the band structures. We mention that if open boundary condition is used for finite systems instead, there will be two Majorana modes at the two ends of the Kitaev  chain~\cite{Kitaev} and two edge states at the ends of the SSH chain~\cite{SSH}.

It is important to examine the dynamic phase more carefully in the limits of zero temperature and infinite temperature.
When $T=0$, $\beta=\infty$. The limit $\tanh(\infty)=1$ leads to  \begin{align}\theta_D(T=0)=-\arctan\left[\tan(\frac{mM\tau}{\hbar})\right]=-\frac{mM\tau}{\hbar}\end{align} 
for the Kitaev chain, and \begin{align}\theta_D(T=0)=\arctan\left[\tan(\frac{J_1\tau}{\hbar})\right]=\frac{J_1\tau}{\hbar}\end{align} 
for the SSH model. The insets of Fig.~\ref{Fig1} confirm the $T=0$ results. If the value of $\frac{M\tau}{\hbar}$ or $\frac{J_1\tau}{\hbar}$ falls outside the range of $\left(2n\pi-\frac{\pi}{2},2n\pi+\frac{\pi}{2}\right)$, $\theta_D(T=0)$ must be carefully evaluated according to Eqs.~(\ref{Ud8}) and (\ref{Ud9}). 

When $T\rightarrow \infty$, $\beta=0$, one can verify that $\theta_D\rightarrow 0$ if $\frac{M\tau}{\hbar}=1.0=\frac{J_1\tau}{\hbar}$ by Eqs.~(\ref{Ud8}) and (\ref{Ud9}). The behavior can also be observed in Fig.~\ref{Fig1} at high temperatures.
However, quantized values will appear if we choose the "resonant values" of $\tau$. If we choose $m\frac{M\tau}{\hbar}=\pm\frac{\pi}{2}=\frac{J_1\tau}{\hbar}$ instead, then
$\theta_D=-\arctan(\pm\infty)=\mp\frac{\pi}{2}$ for the Kitaev chain and $\theta_D=\arctan(\pm\infty)=\pm\frac{\pi}{2}$ for the SSH model since $\tan(\pm\frac{\pi}{2})=\pm\infty$. Interestingly, the values are independent of temperature with the particular choice of $\tau$. 
Moreover, if $m\frac{M\tau}{\hbar}$ and $\frac{J_1\tau}{\hbar}$ are in the range $[-\pi,-\frac{\pi}{2})\cup(\frac{\pi}{2},\pi)$, $\theta_D=\pi$ according to Eqs.~(\ref{Ud8}) and (\ref{Ud9}). (Note $-\pi\equiv\pi$ $\mod 2\pi$.)
Hence, we obtained a surprising result: The dynamic phase at infinitely high temperature is discrete-valued, although it is not a geometric phase and carries no topological information.

\begin{figure}[ht]
\centering
\includegraphics[width=3.2in,clip]
{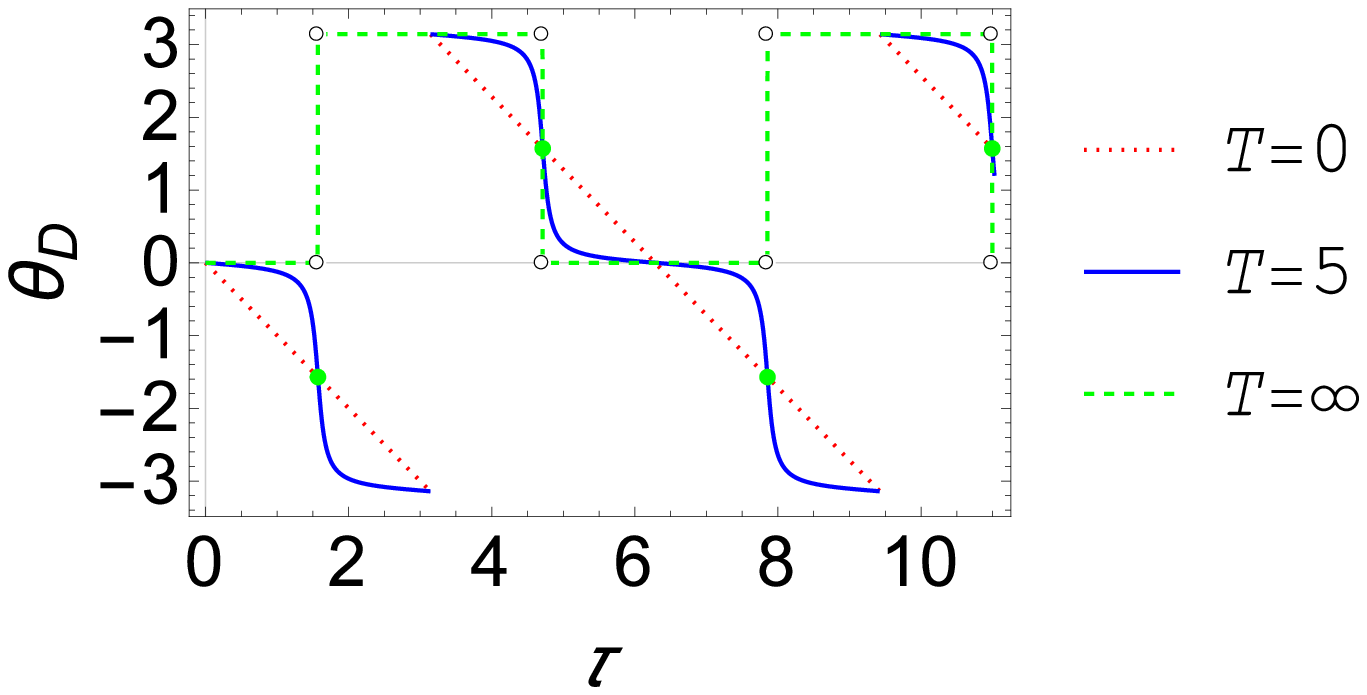}
\includegraphics[width=3.2in,clip]
{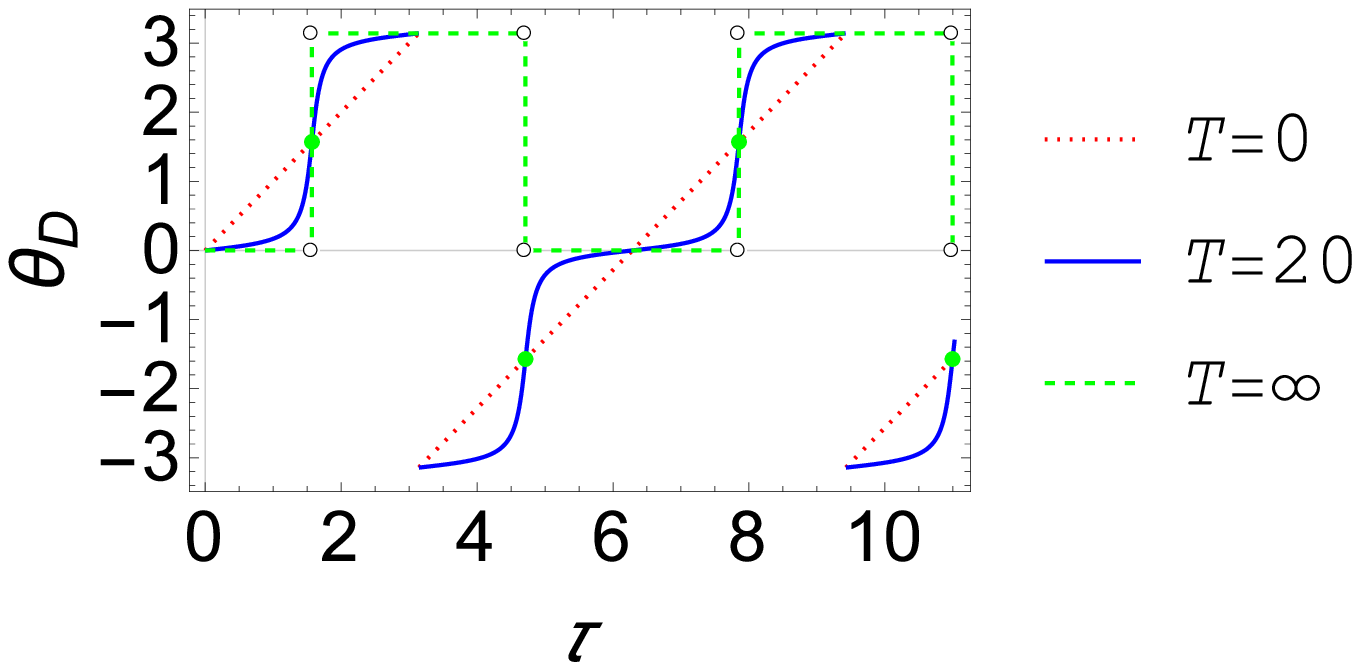}
\caption{(Color online) Dynamic phase as a function of the duration $\tau$ of time-evolution. The top (bottom) panel is for the Kitaev chain with $m=0.6$ and $c=1.0$ (the SSH model with $J_2/J_1=1.2$). $\tau$ is in units of $\hbar/M$ ($\hbar/J_1$) and $T$ is in units of $M/k_B$ ($J_1/k_B$) for the Kitaev chain (SSH model). The dotted, solid, and dashed curves show the behavior at $T=0, 5 ~(20), \infty$ in the respective units. The green solid circles indicate the discrete values that $\theta_D$ takes at infinitely high temperature, and the hollow circles indicate the jumps of $\theta_D$.}
\label{Fig2}
\end{figure}

To explicitly show the quantization of the dynamic phase at infinite temperature, we plot $\theta_D$ as a function of $\tau$ in Fig.~\ref{Fig2} for the two models. The red-dotted, blue-solid and green-dashed lines (green solid circles) are for $T=0, 5 ~(20), \infty$ in units of $\hbar/M$ ($\hbar/J_1$) for the Kitaev chain (SSH model). One can see that $\theta_D$ is a periodic function of $\tau$.
Importantly, $\theta_D$ takes three discrete values at $T=\infty$. For the Kitaev chain, $\theta_D=(-1)^{n-1}\frac{\pi}{2}$ at $\tau=n\pi+\frac{\pi}{2}$ with $n$ being an integer, $\theta_D=0$ if $\tau\in (2n\pi-\frac{\pi}{2},2n\pi+\frac{\pi}{2})$, otherwise $\theta_D=\pi$. For the SSH model, $\theta_D=(-1)^{n}\frac{\pi}{2}$ at $\tau=n\pi+\frac{\pi}{2}$, and the rest results are the same as the those of the Kitaev chain.
When $\tau=2n\pi$, all curves meet at $\theta_D=0$, which is also the root of $\theta(\tau)=0$.
At finite temperatures, $\theta_D$ is a continuous function of $\tau$ within each period.

The point $\tau=n\pi+\frac{\pi}{2}$, or more precisely, $\frac{mM\tau}{\hbar}=n\pi+\frac{\pi}{2}$ for the Kitaev chain and $\frac{J_1\tau}{\hbar}=n\pi+\frac{\pi}{2}$ for the SSH model, should be considered as the ``resonant points", where the values of the dynamic phase are independent of temperature. For convenience, the corresponding values of the dynamic phase will be referred to as the ``resonant values" of $\theta_D$. Away from the resonant points, $\theta_D=0$ at $T=\infty$ as the system becomes totally disordered. The fluctuation of the density of states is overwhelmed by the Boltzmann factor at infinitely high temperature. Thus, the values of the dynamic phase away from the resonant points will be referred to as the ``ordinary value" of $\theta_D$. To confirm that the dynamic phase does not carry topological information, we have tested the two models with other parameters ($m>c$ for the Kitaev chain and $J_2 < J_1$ for the SSH model) and found that the results are qualitatively the same.

\subsubsection{Harmonic Oscillator}\label{ssho}
It is possible that the discrete values of the dynamic phase at infinite temperature may be accidental since what we have considered are simple 1D two-band systems. As a comparison, we consider a system with infinite discrete energy levels, exemplified by the simple harmonic oscillator~\cite{MessiahBook,Merzbacher_book}. The energy levels are $E_n=\hbar\omega(n+\frac{1}{2})$ with $n=0,1,2,3,\cdots,\infty$, where $\omega$ is the angular frequency of the oscillator.
In this situation, the Hamiltonian is independent of time, and we have
\begin{align}\label{Ud4a}
\theta_D&=\arg\text{Tr}[\rho(0)\mathcal{T}\me^{-\frac{\mi}{\hbar} \mathlarger{\oint}\hat{H}\dif t}]\notag\\
&=\arg\left\{\frac{1}{Z}\text{Tr}[\me^{-\beta\hat{H}}\me^{-\frac{\mi}{\hbar}\hat{H}\tau}]\right\}\notag\\
&=\arg\left[\frac{1}{Z}\me^{-\frac{1}{2}(\beta\hbar+\mi\tau)\omega}\sum_{n=0}^\infty\me^{-(\beta\hbar+\mi\tau)n\omega}\right].
\end{align}
Here $Z$ is the canonical partition function.
Since $|\me^{-(\beta\hbar+\mi\tau)\omega}|=\me^{-\beta\hbar \omega}<1$, the geometric series can be evaluated as follows.
\begin{align}\label{Ud4b}
\theta_D&=\arg\left[\frac{1}{Z}\frac{\me^{-\frac{1}{2}(\beta\hbar+\mi\tau)\omega}}{1-\me^{-(\beta\hbar+\mi\tau)\omega}}\right]\notag\\
%&=\arg\left[\frac{1}{Z}\frac{2}{\sinh\frac{\beta\hbar+\mi\tau}{2}\omega}\right]\notag\\
&=\arg\left[\frac{1}{Z}\frac{1}{\sinh\frac{\beta\hbar\omega}{2}\cos\frac{\omega\tau}{2}+\mi\cosh\frac{\beta\hbar\omega}{2}\sin\frac{\omega\tau}{2}}\right].%\notag\\
%&=\arg\left[\frac{1}{Z}\frac{\me^{-\mi\arctan(\tan\frac{\omega\tau}{2}\coth\frac{\beta\hbar\omega}{2})}}{\sqrt{\sinh^2\frac{\beta\hbar\omega}{2}+\sin^2\frac{\omega\tau}{2}}}\right]\notag\\
%&=-\arctan\left(\tan\frac{\omega\tau}{2}\coth\frac{\beta\hbar\omega}{2}\right).
\end{align}
Hence, the exact expressions of the dynamic phase is given by
\begin{widetext}
\begin{align}\label{Ud4b1}
\theta_D=\left\{\begin{array}{cc}
-\arctan\left(\tan\frac{\omega\tau}{2}\coth\frac{\beta\hbar\omega}{2}\right), & \text{if } \frac{\omega\tau}{2} \in 
\left(2n\pi-\frac{\pi}{2},2n\pi+\frac{\pi}{2}\right);\\
\mp\frac{\pi}{2}, & \text{if  } \frac{\omega\tau}{2}=2n\pi\pm\frac{\pi}{2};\\
-\arctan\left(\tan\frac{\omega\tau}{2}\coth\frac{\beta\hbar\omega}{2}\right)-\pi, & \text{if  } \frac{\omega\tau}{2}\in \left(2n\pi+\frac{\pi}{2},2n\pi+\pi\right);\\
-\arctan\left(\tan\frac{\omega\tau}{2}\coth\frac{\beta\hbar\omega}{2}\right)+\pi, & \text{if  } \frac{\omega\tau}{2} \in \big[2n\pi-\pi,2n\pi-\frac{\pi}{2}\big),
\end{array}\right.
\end{align}
\end{widetext}

Interestingly, the dynamic phase also exhibits four discrete values at $T=\infty$, as explained below. If $\omega\tau=2n\pi$ for some integer $n$, $\theta_D=\arg(\cos(n\pi))=\arg((-1)^n)=\frac{1-(-1)^n}{2}\pi$, independent of temperature. If $\omega\tau\neq 2n\pi$ and $T=\infty$,
\begin{align}\label{tdho}\theta_D=-\arg(\mi\sin\frac{\omega\tau}{2})=-\text{sgn}(\sin\frac{\omega\tau}{2})\frac{\pi}{2}
\end{align}
since $\sinh(0)=0$ and $\cosh(0)=1$.
For the harmonic oscillator at $T=\infty$, therefore, we have $\theta_D=0$ when $\omega\tau=0$, $\theta_D=\pi$ when $\omega\tau=2\pi$,  $\theta_D=-\frac{\pi}{2}$ when $\omega\tau \in (0,2\pi)$, and $\theta_D=\frac{\pi}{2}$ when $\omega\tau \in (2\pi,4\pi)$ in the range $[0,4\pi)$.
%We mention that at $T=0$, $\theta_D=-\frac{\omega\tau}{2}$ since $\coth({\infty})=1$.

\begin{figure}[ht]
\centering
\includegraphics[width=2.7in,clip]
{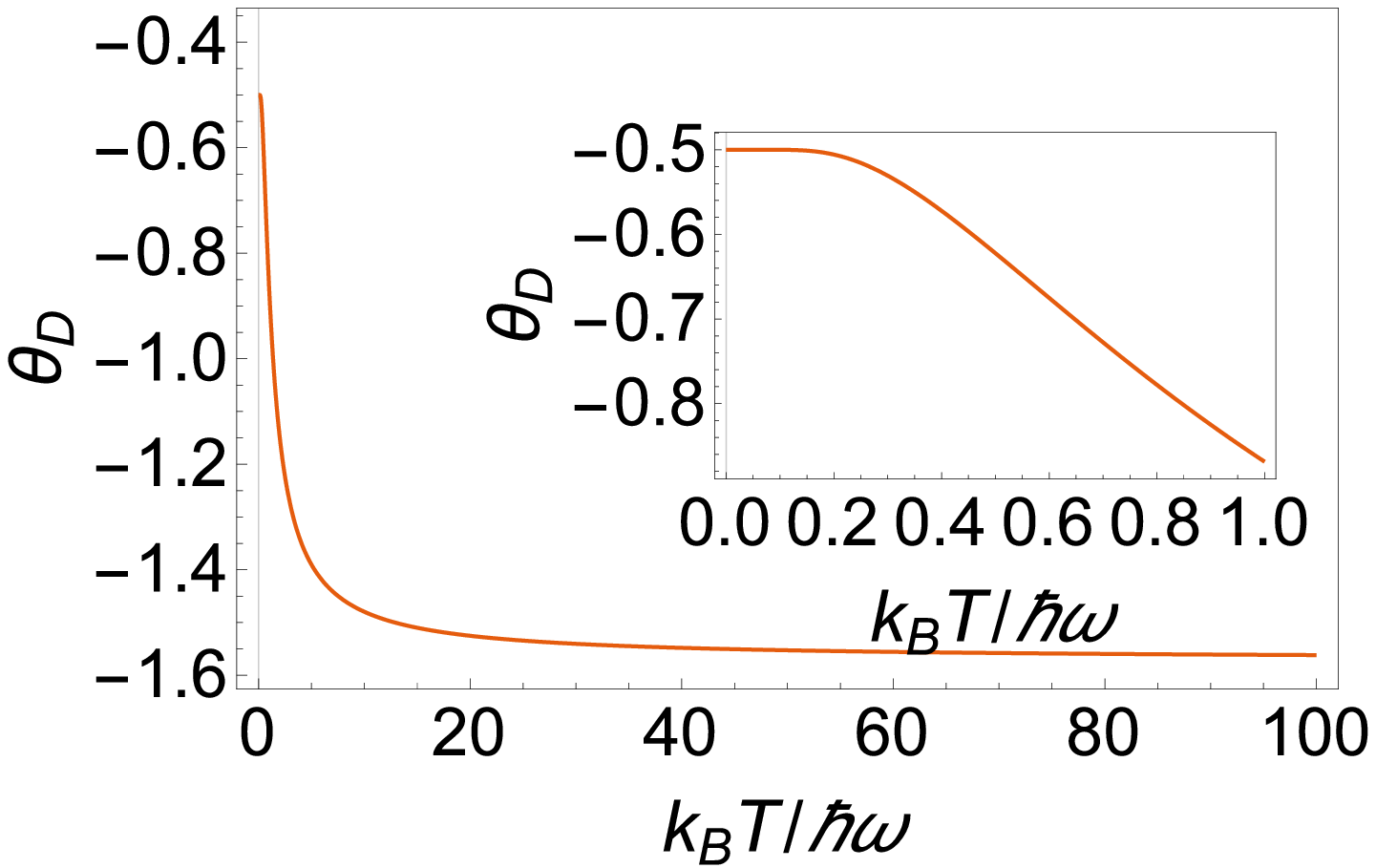}
\includegraphics[width=3.3in,clip]
{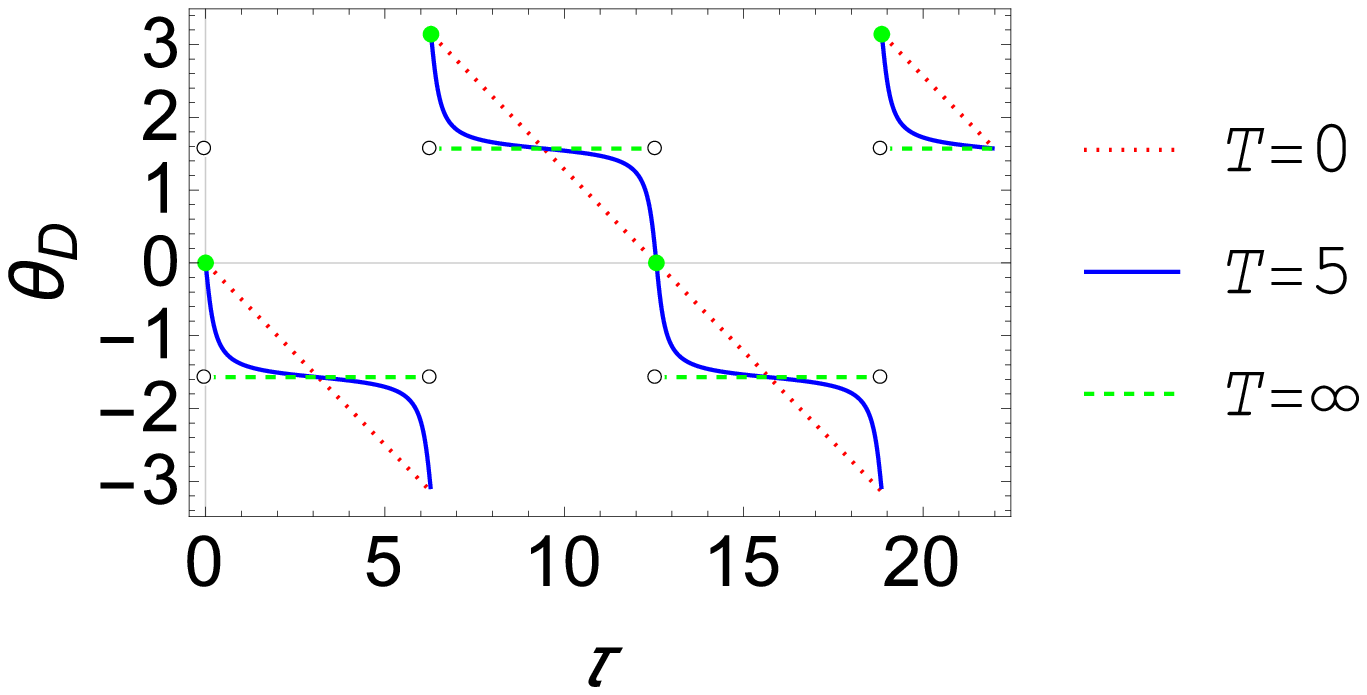}
\caption{(Color online). The top panel shows the behavior of $\theta_D$ as a function of temperature (in units of $\hbar\omega/k_B$) for the harmonic oscillator, where the parameter is chosen as $\omega\tau=1.0$. The bottom panel plots $\theta_D$ vs. $\tau$ (in unites of $1/\omega$) at $T=0,5,\infty$ (in units of $\hbar\omega/k_B$) for the harmonic oscillator in dotted, solid, and dashed lines, respectively. }
\label{Fig3}
\end{figure}

The dynamic phase of a harmonic oscillator as a function of temperature with a selected period $\omega\tau=1.0$ and as a function of the period $\tau$ at selected temperatures is shown in Figure~\ref{Fig3}.  For the top panel, the chosen value of $\omega\tau=1$, $\theta_D$ exhibits the ``ordinary values" at finite temperatures, showing a smooth curve. The inset magnifies the detail at low temperatures, and we confirm that $\theta_D=-\frac{\omega\tau}{2}$ at $T=0$ if $\omega\tau=1.0$. 

To investigate the $\tau$ dependence and compare with Fig.~\ref{Fig2}, the lower panel of Figure~\ref{Fig3} shows $\theta_D$ as a function of $\tau$ at $T=0$, $5$, and $\infty$ (in units of $\hbar\omega/k_B$). The four discrete values at infinite temperature can be clearly spotted. A detailed comparison of Figs.~\ref{Fig2} and \ref{Fig3} shows that, for both the harmonic oscillator and the two-band models, the ``resonant values" of $\theta_D$ at $T=\infty$ are quantized and take two discrete values, but the two values differ in the different systems. Interestingly, the ``ordinary valuess" at infinite temperature are also quantized and take two discrete values as well. The four values exhaust the list of possibilities of $\theta_D$ at infinite temperature for the three examples presented here.

\subsubsection{Continuous energy spectrum without band gap}
Furthermore, we consider the limit where the energy-level spacing vanishes, approaching a continuous energy spectrum without a band gap. In this situation, the dynamic phase becomes
\begin{align}\label{Ud4c}
\theta_D&=\arg(\sum_{n=1}^N\frac{1}{Z}\me^{-\frac{E_n}{\hbar}(\beta\hbar+\mi\tau)}) \notag \\
&\xrightarrow{N\rightarrow\infty}\arg\left(\frac{1}{Z}\int_{0}^\infty\dif E\me^{-\frac{E}{\hbar}(\beta\hbar+\mi\tau)}\right).
\end{align}
Here the range of the energy is from $0$ to $\infty$, and $\langle n|\rho(0)|n\rangle=\frac{1}{Z}\me^{-\beta E_n}$ is the density of states at $E_n$. Explicitly,
\begin{align}\label{Ud4d}
\theta_D=\arg\left(\frac{\hbar}{Z}\frac{1}{\beta\hbar+\mi\tau}\right)=-\arctan(\frac{\tau}{\beta\hbar}).
\end{align}
When $T\rightarrow \infty$ ($\beta\rightarrow 0$), we have
\begin{align}\label{Ud4e}
\theta_D\rightarrow-\frac{\pi}{2},
\end{align}
regardless of $\tau$. If $E_0\neq 0$, one may define a Hamiltonian $\hat{H}^{\prime}=\hat{H}-E_0 1$, where $1$ is the identity operator, and the result remains the same as if $E_0=0$.

The difference between the case with a continuous spectrum and the former examples lies in the existence of a discrete $\Delta E$, which may be from the discrete energy-levels or a band gap. For example, $\Delta E=\hbar \omega$ for the harmonic oscillator. As a consequence, one of the $\theta_D$ occurs at the ``resonant point" when $\frac{1}{\hbar}\Delta E\tau=\omega\tau=2n\pi$ with an integer $n$. Similar arguments apply to the two-band models with a finite band gap causing the resonant points. In the case with a continuous spectrum, such resonant points do not exist for any finite $\tau$. In other words, the dynamic phase fails to acquire resonant values in the continuum limit without an energy gap. Moreover, even the ``ordinary value" of the dynamic phase is uniquely fixed according to Eq.~(\ref{Ud4d}). For the harmonic oscillator, the signs of the ``ordinary values" are determined by $\tan(\frac{\omega\tau}{2})$ according to Eq.~(\ref{tdho}). The procedure becomes ill-defined when $\omega\rightarrow 0$ (or $\Delta E\rightarrow 0$). Hence, the dynamic phase at infinite temperature offers an indication of the existence of an energy spacing from a discrete spectrum or band gap if it takes multiple discrete values.

\subsection{Implications for classical systems}
One may be interested in the counterpart of  the dynamic phase $\theta_D$ for classical mixed states. However, the concept of the mixed state in classical mechanics is only sparsely explored in the literature~\cite{AJP65,Res12,JGM19}, and it seems there is no broadly accepted definition. For example, the dynamic phase of a classical object following a closed curve in the parameter space was introduced in Ref.~\cite{GPbook}. However, the objects discussed there have definite and traceable trajectories, so they should be more appropriately viewed as classical pure states.

Here, we adopt the idea discussed in Ref.~\cite{Res12}. The mixed states in classical systems may be identified as the statistical ensembles. Mathematically, they can be described by the probability density in the classical phase space~\cite{Res12}, which is the counterpart of the density matrices in quantum mechanics. The associated equation of motion is described by the Liouville equation. However, there is no procedure in classical mechanics acting like the purification of the density matrix in quantum mechanics. Thus, what we have done for defining the dynamic phase of mixed quantum states cannot be applied to classical systems directly, so a well-defined dynamic phase for mixed classical states remains elusive.

Nevertheless, there might be another option. Since every mixed classical state is associated with a probability, the corresponding dynamic phase may simply be defined as the weighted summation of those from the constituent processes, similar to Eq.~(\ref{t1}). However, the classical expression, if so defined, differs from the quantum dynamic phase, which is the phase of a coherent sum of the constituents. Therefore, the dynamical phase may serve as an indicator for distinguishing the mixed states from quantum or classical systems.

\section{Conclusion}\label{Conclusion}
By using the fiber-bundle language, we have shown that the Berry process is compatible with the Schrodinger equation and the dynamics of a pure quantum state can be described by an effective adiabatic process. In contrast, the Ulhmann process of a mixed quantum state is generally not compatible with the time-evolution equation of the density matrix according to the Hamiltonian. A general mixed quantum state then accumulates a dynamic phase during its time evolution. For cyclic and quasi-static processes, the dynamic phase may take multiple discrete values at infinite temperature, as shown by the examples of the 1D two-band models and the harmonic oscillator, but the behavior is absent for a system with a continuous energy spectrum and no band gap. Although the dynamic phase does not reveal geometric information, it reflects the underlying energy spectrum and may accompany the genuine topological phase that may be discovered in future research.

\begin{acknowledgments}
H. G. was supported by the National Natural Science Foundation of China (Grant No. 11674051).
\end{acknowledgments}

\appendix
\section{Details of Berry phase}\label{appBerry}

\subsection{Berry bundle and Berry phase}\label{appBerry1}
Let $M$ be the parameter space that may be considered as a manifold described by the local coordinates $\mathbf{R}$.
The physical state is normalizable, hence we define $\mathbb{P}:=\{|\mathbf{R}\rangle\big|\langle \mathbf{R}|\mathbf{R}\rangle=1\}$. The proper quantum phase space of this system is the space of rays given by $H=\mathbb{P}/\sim$ since a quantum state $|\mathbf{R}\rangle$ cannot be distinguished from the state $\me^{\mi\theta}|\mathbf{R}\rangle$ where $\me^{\mi\theta}\in$ U(1). More generally, two states $|\psi\rangle$, $|\phi\rangle\in H$ are physically equivalent if $|\psi\rangle=c|\phi\rangle$ if $c$ is a complex number (Note $H$ is distinct from $M$ though a point $|\bR\rangle\in H$ is parameterized by the local coordinate $\bR\in M$).

Thus, we can define a U(1)-principle bundle $P(H,\textrm{U(1)})$ where $H$ is the base manifold, and
the fibre $F_\bR$ at each point $|\bR\rangle$ of $H$ consists of the equivalent class of quantum states
\begin{align}
\pi^{-1}(|\mathbf{R}\rangle)=[|\mathbf{R}\rangle]\equiv\{g|\mathbf{R}\rangle|g\in \textrm{U(1)}\}.
\end{align}
Here $\pi$ is the projection, which satisfies
\begin{align}
\pi\circ\phi(|\mathbf{R}\rangle,g|\mathbf{R}\rangle)=|\mathbf{R}\rangle;\quad  \forall g\in \textrm{U(1)},
\end{align}
where $\phi$ is a local trivialization.
$U(1)$ is the structure group, which is isomorphic to the fiber and acts on the fiber from the right, i.e.,
\begin{align}
\phi^{-1}(|\mathbf{R}\rangle )g=(|\mathbf{R}\rangle,g'|\mathbf{R}\rangle)g=(|\mathbf{R}\rangle,g'g|\mathbf{R}\rangle).
\end{align}
A section $\sigma: H\rightarrow P$ is a smooth map which satisfies $\pi\circ \sigma=1_H$. Locally fixing the phase of $|\bR\rangle$ at each point $|\mathbf{R}\rangle\in H$ amounts to choosing a specific section (wave-function) $\sigma(|\bR\rangle)=\me^{\mi\theta(\bR)}|\bR\rangle$. Obviously, we can construct a Hermitian scalar product on $P$
\begin{align}\label{hs}
h(|\mathbf{R}_1\rangle,|\mathbf{R}_2\rangle):=\langle \mathbf{R}_1|\mathbf{R}_2\rangle.
\end{align}

For two arbitrary vector fields $X$ and $Y$, it can be shown that
\begin{align}\label{FXY}
F(X,Y)&=\langle X(\mathbf{R})|Y(\mathbf{R})\rangle-\langle Y(\mathbf{R})|X(\mathbf{R})\rangle\notag\\&=2\mi\textrm{Im}\langle X(\mathbf{R})|Y(\mathbf{R})\rangle.
\end{align}
Hence, $\frac{1}{2\mi}F$ is the imaginary part of the hermitian structure defined by Eq.~(\ref{hs}), which is also a symplectic structure on $P$.

\subsection{Connection and Horizontal and Vertical Subspaces}\label{appBerry2}
Let the loop $\gamma$ be parameterized by $t$, and $X$ be the tangent vector to $\gamma(t)$. Since the local coordinate of $|\bR\rangle$ can be expressed as $\bR=(R_1,R_2,\cdots,R_k)$, then $X$ can be locally expressed as
\begin{align}\label{X}
X=X^i\frac{\partial }{\partial R^i}=\frac{\dif R^i}{\dif t}\frac{\partial }{\partial R^i}=\frac{\dif }{\dif t}.
\end{align}
Note $X\in TM\cong M$, assuming the base vectors of $M$ are $\mathbf{e}_1,\cdots,\mathbf{e}_k$, then the tangent operator $X$ can also be expressed as
\begin{align}\label{tX}
\mathbf{X}\equiv X^i\mathbf{e}_i=\frac{\dif R^i}{\dif t}\mathbf{e}_i=\frac{\dif }{\dif t}(R^i\mathbf{e}_i)=\frac{\dif }{\dif t}\mathbf{R}=X(\mathbf{R}),
\end{align}
where we have assumed that $M$ is locally flat.
In general, one has
$
\tilde{\gamma}(1)=\tilde{\gamma}(0)g_\gamma(1)$,
where the element $g_\gamma\in $U(1) defines a transformation (associated with the loop $\gamma$) on the fiber. The set of $g_\gamma$ forms a subgroup of the structure group U(1), which is called the holonomy group. For convenience, we call it the Berry holonomy hereafter.

Without loss of generality, we assume $\gamma(0)=|\mathbf{R}\rangle$, $\gamma(t)=|\mathbf{R}(t)\rangle$. Since $\mathbf{X}$ is the tangent direction of $\mathbf{R}(t\mathbf{})$, then $|\mathbf{R}(t)\rangle=|\mathbf{R}+t\mathbf{X}\rangle$ up to the first order of $t$ if $t$ is is infinitesimally small. The horizontal lift  $\tilde{\gamma}(t)$ defines a section, which is a  wave-function:
\begin{align}\label{wf}
\tilde{\gamma}(t)\equiv \sigma(|\mathbf{R}(t)\rangle)=\me^{\mi\theta(t)}|\mathbf{R}(t)\rangle.
\end{align}
Obviously $\theta(0)=0$.
Let $\tilde{X}$ be the tangent vector field to $\tilde{\gamma}(t)$, then $\tilde{X}\in TP$ where $TP$ is the tangent bundle associated with $P$.
The fact $\pi\circ\tilde{\gamma}=\gamma$ leads to
\begin{align}\label{wf2}
\pi_*\tilde{X}=X.
\end{align}
Since $\tilde{X}$ is tangent to the horizontal lift of $\gamma(t)$, then it must be a horizontal vector.

Here we need to define the ``horizontal vector'' more precisely.
The tangent bundle $TP$ can be separated into the horizontal ($HP$) and vertical ($VP$) subspaces as $TP=HP\oplus VP$,
hence $\tilde{X}\in HP$.
The vertical space is the subspace in which all vectors are tangent to the fibre.
At a point $|\mathbf{R}\rangle \in H$, the fibre is $\pi^{-1}(|\mathbf{R}\rangle)=\{\me^{\mi \theta}|\mathbf{R}\rangle\big|\theta\in \mathds{R}\}$. Therefore, a vector in the vertical subspace at $|\mathbf{R}\rangle$ can be given by
\begin{align}
\frac{\dif }{\dif s}(\me^{\mi\theta s}|\mathbf{R}\rangle)\Big|_{s=0}=\mi\theta|\mathbf{R}\rangle.
\end{align}
In other words, we have
\begin{align}\label{VPR}
VP_{|\mathbf{R}\rangle}=\{\mi\theta|\mathbf{R}\rangle\big|\theta\in \mathds{R}\},
\end{align}
which is equivalent to u(1)$\cong\mi \mathds{R}$.
Obviously, Eq.(\ref{VPR}) indicates that any point $|\phi\rangle\in \pi^{-1}(|\mathbf{R}\rangle)$ must satisfy the fact that $\langle\phi|\mathbf{R}\rangle$ is a unit complex number, and $\mi\arg\langle\phi|\mathbf{R}\rangle\in VP_{|\mathbf{R}\rangle}$. Note $\arg\langle\phi|\mathbf{R}\rangle$ is ill-defined if $\langle\phi|\mathbf{R}\rangle=0$.
On the contrary, if a quantum state vector $|\mathbf{R}'\rangle$ satisfies $\langle\mathbf{R}'|\mathbf{R}\rangle\neq 0$, then $\mi\arg\langle\Phi|\mathbf{R}\rangle\in VP_{|\mathbf{R}\rangle}$. That is to say, $|\mathbf{R}'\rangle$ must contain a component which belongs to $VP_{|\mathbf{R}\rangle}$. A horizontal vector must not contain any component in the vertical subspace. Thus, the horizontal subspace at $|\mathbf{R}\rangle$ can be defined as
\begin{align}\label{HPR}
HP_{|\mathbf{R}\rangle}=\{|\psi\rangle\big|\langle \psi|\mathbf{R}\rangle=0\}.
\end{align}

For convenience, let \begin{align}\label{psit0}|\psi(t)\rangle=\me^{\mi\theta(t)}|\mathbf{R}(t)\rangle\end{align} denote a point on the curve $\tilde{\gamma}$. Similarly, the tangent vector at $|\mathbf{R}\rangle$ is given by
\begin{align}\label{psit}
\tilde{X}|\psi(t)\rangle\equiv \frac{\dif_P|\psi(t)\rangle }{\dif t}\Big|_{t=0}.
\end{align}
The horizontality condition (\ref{IHC0}) can also be expressed as
\begin{align}\label{CPHC}
\omega_{|\psi\rangle}(\tilde{X})=0,
\end{align}
i.e. the horizontal vector $\tilde{X}$ belongs to the kernel of $\omega$.

\subsection{Berry phase from the connection}\label{appBerry3}
The connection $\omega$ is actually a projection of $TP$ onto $HP$. In Ref.~\cite{Nakahara}, Eq.~(\ref{CPHC}) is instead applied as the definition of the horizontal subspace.
The pull-back of $\omega$ by the section defined in Eq.~(\ref{wf}) introduces a connection on $H$
\begin{align}\label{BA}A_B=\sigma^*\omega.\end{align}
Here we include a subscript ``$B$'' because $A_B$ is in fact the Berry connection. Plugging in Eqs.~(\ref{wf}) and (\ref{CP}), the definition (\ref{BA}) leads to
\begin{align}\label{Bc0}
A_{B}=\langle \mathbf{R}|\dif|\mathbf{R}\rangle\quad \text{or}\quad \langle \mathbf{R}|\dif\mathbf{R}\rangle.
\end{align}
In the component form, it is
\begin{align}\label{Bc2}
A_{Bi}=\langle \mathbf{R}|\frac{\partial}{\partial R^i}|\mathbf{R}\rangle.
\end{align}

Let $g_\gamma(t)=\me^{\mi\theta(t)}$, then Eq.~(\ref{OmegaX}) further reduces to
\begin{align}\label{OmegaX1}
A_B(X)+g_\gamma(t)^{-1}\frac{\dif g_\gamma(t)}{\dif t}=0
\end{align}
or
\begin{align}\label{ox}
\frac{\dif g_\gamma(t)}{\dif t}=-A_B(X)g_\gamma(t),
\end{align}
subject to the condition $g_\gamma(0)=1$.
If we express $g_\gamma(t)^{-1}\frac{\dif g_\gamma(t)}{\dif t}=g_\gamma(t)^{-1}\dif g_\gamma(\tilde{X})$, then Eq.~(\ref{OmegaX1}) implies that $\omega$ given by Eq.~(\ref{CP}) is also expressed as
\begin{align}\label{ox2}
\omega=\pi^*A_B+g_\gamma^{-1}\dif g_\gamma.
\end{align}
which satisfies Eq.~(\ref{BA}). In Ref.~\cite{Nakahara}, Eq.~(\ref{ox2}) was used instead to construct a connection over the total space from a connection on the base manifold.
The formal solution to Eq.~(\ref{ox}) is
\begin{align}\label{sox}
g_\gamma(t)&=\mathcal{P}\me^{-\int_0^tA_{Bi}\frac{\dif R^i}{\dif \tau}\dif \tau}=\mathcal{P}\me^{-\int_{\gamma(0)}^{\gamma(t)}A_B(X(\tau))\dif \tau} \notag \\
&=\mathcal{P}\me^{-\int_0^tA_B},
\end{align}
where $\mathcal{P}$ is a path-ordering operator along $\gamma(t)$. Because $U(1)$ is an abelian group, so is the Berry holonomy. Therefore, the operator $\mathcal{P}$ can be safely omitted, and we obtain
\begin{align}\label{sox1}
g_\gamma(t)=\me^{-\int_0^tA_B(X(\tau))\dif \tau}.
\end{align}
When $t=1$, $\gamma(0)=\gamma(1)$ we get the holonomy element
\begin{align}\label{sox2}
g_\gamma(1)\equiv\me^{\mi\theta_B}=\me^{-\oint A_B}.
\end{align}
From the expression we obtain the Berry phase described in Sec.~\ref{CB}.

The Berry curvature is a two-form defined by $F=\dif A_B$ By Eq.~(\ref{Bc0}), we have
\begin{align}\label{FB}
F=\langle\dif \mathbf{R}|\wedge|\dif \mathbf{R}\rangle
\end{align}
where $\wedge$ is the wedge operator.

As discussed in the main text, two nonzero states are equivalent if one is a scalar multiplication of another. This relation is reflexive, symmetric and transitive. The resulting space of equivalence classes is the projective Hilbert space $H$, which naturally carries a metric structure known as the Fubini-Study metric~\cite{Uhlmann86}. It is given in terms of the Hilbert space distance
\begin{align}\label{dfs}
\dif^2_\textrm{FS}([\psi_1],[\psi_2]):=\inf |||\psi_1\rangle-|\psi_2\rangle||,
\end{align}
where the infimum is taken over all normalized representatives $|\psi_i\rangle$ of equivalence classes. One can show that
$\dif^2_\textrm{FS}([\psi_1],[\psi_2])
%&=\inf (\langle\psi_1|-\langle\psi_2|)(|\psi_1\rangle-|\psi_2\rangle)\notag\\
%&=2-\sup(\langle\psi_1|\psi_2\rangle+\langle\psi_2|\psi_1\rangle)\notag\\
=2-2\sup\textrm{Re}(\langle\psi_1|\psi_2\rangle)
\le2-2|\langle\psi_1|\psi_2\rangle|$.
Here the supremum of $\textrm{Re}(\langle\psi_1|\psi_2\rangle)$ is realized if $\textrm{Im}(\langle\psi_1|\psi_2\rangle)=0$, i.e. $\langle\psi_1|\psi_2\rangle$ is a positive real number. If we let $|\psi_2\rangle\rightarrow |\psi_1\rangle$ along a curve $\tilde{\gamma}(t)$, this condition further reduces to $\textrm{Im}(\langle\psi_1| \dot{\psi}_1\rangle)=0$, which is exactly the condition for a parallel transport of $|\psi_1\rangle$ along $\tilde{\gamma}(t)$ to $|\psi_2\rangle$ according to Eq.~(\ref{IHC}).
Hence the infimum (\ref{dfs}) is realized if the representatives satisfies Eq.~\eqref{p1}.

\section{Details of Uhlmann phase}\label{appUhlmann}
\subsection{Purification of Density Matrix}\label{appUhlmann1}
Assuming the dimension of the Hilbert space formed by the mixed quantum states considered here is $n$,
there is a U$(n)$-gauge degrees of freedom in the choice of the amplitude: Both $W$ and $WV$ are amplitudes of the same density matrix $\rho$ if $V\in$ U$(n)$ is an arbitrary unitary operator. Explicitly,
\begin{align}
\rho=WW^\dagger=(WV)(WV)^\dagger.
\end{align}
The amplitude plays the role of $|\psi\rangle$ in the discussion of the Berry phase. Moreover, the amplitudes also
form a Hilbert space $H_W$, where a scalar product, called the Hilbert-Schmidt product, is defined as
\begin{align}\label{HSip}
(W_1,W_2):=\textrm{Tr}(W^\dagger_1W_2).
\end{align}
It can be verified that it is also a Hermitian scalar product~\cite{GPbook}.

One may express the density matrix in the space $\mathcal{H}$ spanned by its eigenvectors as
\begin{align}\label{r1}
\rho=\sum_i\lambda_i|i\rangle\langle i|,
\end{align}
then the amplitude associated with it is
\begin{align}\label{w1}
W=\sum_i\sqrt{\lambda_i}|i\rangle\langle i|U,
\end{align}
where $U$ is the phase factor of the corresponding density matrix.
The purification is in fact an isomorphism between the spaces $H_W$ and $\mathcal{H}\otimes\mathcal{H}$:
\begin{align}\label{w2}
W=\sum_i\sqrt{\lambda_i}|i\rangle\langle i|U\leftrightarrow|W\rangle=\sum_i\sqrt{\lambda_i}|i\rangle\otimes U^T|i\rangle,
\end{align}
where $U^T$ is the transportation of $U$ taken with respect to the eigenbasis of $\rho$.
Hence by taking the partial trace over the second Hilbert space of $\mathcal{H}\otimes\mathcal{H}$ we can obtain the density matrix
\begin{align}\label{dme1}
\rho=\textrm{Tr}_2(|W\rangle\langle W|),
\end{align}
where $\textrm{Tr}_2$ is the partial trace taken over the second Hilbert space of $\mathcal{H}\otimes\mathcal{H}$.
Moreover, it can be shown that the inner product between two pure states gives the Hilbert-Schmidt product
\begin{align}\label{ipe1}
\langle W_1|W_2\rangle=\textrm{Tr}(W^\dagger_1W_2).
\end{align}
The left-hand-side is a structure of the Hermitian scalar product defined on $\mathcal{H}\otimes\mathcal{H}$.

\subsection{Connection and the Horizontal and Vertical Subspaces}\label{appUhlmann2}
Let $\gamma:[0,1]\rightarrow M$ be a loop in $M$, which induces a loop in $Q$ defined by $\rho(\gamma(t))\equiv \rho(\mathbf{R}(t))$. Here $Q$ is the space spanned by the density matrix.
The horizontal lift $\tilde{\gamma}$ is a curve in $E$, which introduces the Uhlmann holonomy illustrated as follows. Let
 \begin{align}
 \tilde{\gamma}(0)=\sqrt{\rho(\gamma(0))}V(\gamma(0)),\quad  \tilde{\gamma}(1)=\sqrt{\rho(\gamma(1))}V(\gamma(1)),
\end{align}
where $\rho(\gamma(0))=\rho(\gamma(1))$ since $\gamma(0)=\gamma(1)$. However, $\tilde{\gamma}$ may not be a closed curve, then $V(\gamma(0))$ may be different from $V(\gamma(1))$. They are off by an element of the Uhlmann holonomy
\begin{align}
V(\gamma(1))=V(\gamma(0))g_\gamma(1).
\end{align}

The tangent vector of $\tilde{\gamma}$ belongs to the tangent bundle $TE$. Since $\tilde{\gamma}$ is a horizontal lift of $\gamma$, its tangent vector must belong to the horizontal subspace of $TE$, which is denoted by $HE$. Moreover, $HE$ is the complement of the vertical subspace $VE$ of $TE$. Here $TE$ is the tangent space of a fibre space.
In other words, the tangent bundle can be separated as $TE=HE\oplus VE$
At the point $W$, it can be shown that the vertical and horizontal subspaces are given by
\begin{align}
VE_W &=\{Y\in T_WE|Y^\dagger W+W^\dagger Y=0\},\notag \\ HE_W &=\{Y\in T_WE|Y^\dagger W-W^\dagger Y=0\}.
\end{align}
The details are summarized below, along with the parallel transport condition expressed in terms of the decomposition.
Let $\tilde{X}$ be the tangent vector of the curve $\tilde{\gamma}(t)$. Since $\tilde{\gamma}(t)$ is a horizontal lift of the loop $\gamma(t)$, then $\tilde{X}$ must belong to the horizontal subspace of the tangent bundle $TE$.

A curve lying in the fiber $F_W$ at the point $W$ can be expressed as $WU(s)$ since $\pi(WU(s))=\pi(W)$, where $U(s)=\me^{su}\in $ U$(n)$ with $s$ being the parameter.
The generator $u$ must be a $n$-dimensional anti-hermitian matrix, i.e.
\begin{align}\label{uah}
u=-u^\dagger.
\end{align}
The tangent vector of $WU(s)$ must be a vertical vector, which by definition belongs to $VE$:
\begin{align}\label{Vt}
\dot{W}_V\equiv\frac{\dif }{\dif s}WU(s)|_{s=0}=Wu.
\end{align}
Since $\dot{W}_V\in V_E$, i.e it is the vertical part of $\dot{W}$, then $\pi_* \dot{W}_V=0$, which can be directly verified by using $\pi(WU(s))=WU(s)(WU(s))^\dagger =WW^\dagger $
\begin{align}\label{Vt2}
\pi_* \dot{W}_V&=\frac{\dif }{\dif s}\big[W\me^{su}(W\me^{su})^\dagger\big]\Big|_{s=0}\notag\\&=WuW^\dagger+Wu^\dagger W^\dagger\notag\\&=0,
\end{align}
where Eq.~(\ref{uah}) has been applied.
This can also be written as
\begin{align}\label{Vt3}
W\dot{W}_V+\dot{W}_VW=0.
\end{align}

Let $\tilde{\gamma}(t)=W(t)$. Along the horizontal flow which is directed by $\tilde{X}$, we have
\begin{align}
W(t+\dif t)=W(t)+\dif t \tilde{X}(W),\text{ i.e. } \tilde{X}(W)=\dot{W}_H
\end{align}
where the subscript `$H$' means a horizontal direction. Since $\tilde{X}$ must be perpendicular to any tangent direction of a fiber, then a generalization to Eq.~(\ref{IHC0}) gives
\begin{align}\label{upcd1}
\langle \tilde{X}(W)|Wu\rangle=\text{Tr}(\dot{W}_H^\dagger W u)=0, \quad \forall u \in \text{u(n)}.
\end{align}
Taking the Hermitian conjugate of this condition, we get
\begin{align}\label{upcd2}
0=\textrm{Tr}(u^\dagger W_H^\dagger\dot{W})=-\textrm{Tr}(uW^\dagger\dot{W}_H),
\end{align}
where Eq.~(\ref{uah}) has been applied. Combining Eqs.~(\ref{upcd1}) and (\ref{upcd2}), we get
\begin{align}\label{upcd3}
0=\textrm{Tr}\big[u(\dot{W}_H^\dagger W-W^\dagger\dot{W}_H)\big].
\end{align}
This is true for any anti-Hermitian matrix $u$. Hence, the parallel transport condition is satisfied if
\begin{align}\label{upcd4}
\dot{W}_H^\dagger W=W^\dagger\dot{W}_H.
\end{align}

\subsection{Uhlmann Connection, Uhlmann Curvature and Uhlmann Phase}\label{appUhlmann3}
Note $\dot{W}_H=\dif W(\tilde{X})$ where $\dif W$ is a one-form, then Eq.(\ref{upcd4}) can be reexpressed as
\begin{align}\label{UC1}
W^\dagger\dif W(\tilde{X})-\dif W^\dagger(\tilde{X})W=0.
\end{align}
Similar to Eq.(\ref{CP}), it seems that a natural choice of $\omega$ is
\begin{align}\label{UC2}
\omega=W^\dagger\dif W-\dif W^\dagger W
\end{align}
which is a u$(n)$-valued one-form. However, Eq.~(\ref{UC2}) is not a proper definition since it does not transform properly under the gauge transformation $W\rightarrow WV$ where $V \in \text{U}(n)$. In contrast, the expression \eqref{UC3} transforms properly under the gauge transformation $W'=WV$.

The connection $\omega$ shown in Eq.~\eqref{UC3} still satisfies Eq.~(\ref{OmegaXU}) for $\tilde{X}\in HE$. Furthermore, it can be verified that if $Y\in VE$, \begin{align}\label{oY}
\omega(Y)\in \text{u}(n).
\end{align}
Eq.~(\ref{UC3}) is an implicit equation for $\omega$. We are interested in the Uhlmann connection, which is defined by \begin{align}\label{AU}A_U=\sigma^*\omega\end{align} on the base space $Q$. Here $\sigma$ is the section $\sigma(\rho)=W$. To find the expression of the Uhlmann connection, we substitute $W=\sqrt{\rho}U$ into both sides of Eq.~(\ref{UC3}) and get
\begin{align}\label{UC4}
&U^\dagger[\sqrt{\rho},\dif\sqrt{\rho}]U+U^\dagger\rho\dif U+U^\dagger \dif U U^\dagger \rho U \notag \\
&=U^\dagger \rho U\omega+\omega U^\dagger \rho U,
\end{align}
where we have applied $\dif U^\dagger=-U^\dagger \dif U U^\dagger$. A possible construction of $\omega$ that satisfies Eqs.~(\ref{OmegaXU}), (\ref{oY}), and (\ref{AU}) is~\cite{Nakahara}
\begin{align}\label{AU4}
\omega=U^\dagger \pi^*A_UU+U^\dagger\dif U,
\end{align}
which is also a generalization of Eq.~(\ref{ox2}).
Substitute Eq.~(\ref{AU4}) into Eq.~(\ref{UC4}), we get
\begin{align}\label{AU5}
\rho \pi^*A_U+\pi^*A_U\rho=[\sqrt{\rho},\dif\sqrt{\rho}].
\end{align}
For a horizontal vector $\tilde{X}$, $\pi^*A_U(\tilde{X})=A_U(\pi_*\tilde{X})=A_U(X)$ where $X=\pi_*\tilde{X}\in TQ$ (see Eq.~(\ref{wf2}) for the Berry bundle). Then, evaluating Eq.~(\ref{AU5}) on an arbitrary $\tilde{X}$, we get
\begin{align}\label{AU6}
\rho A_U(X)+A_U(X)\rho=[\sqrt{\rho},\dif\sqrt{\rho}(\tilde{X})].
\end{align}
Similarly, evaluating both sides of Eq.~(\ref{UC4}) and applying Eq.~(\ref{OmegaXU}), we get
\begin{align}\label{AU7}
\rho\dif U(\tilde{X})U^\dagger+\dif U(\tilde{X})U^\dagger\rho=-[\sqrt{\rho},\dif\sqrt{\rho}(\tilde{X})].
\end{align}
Comparing Eqs.~(\ref{AU6}) and (\ref{AU7}), we get
\begin{align}\label{AU8}
A_U(X)=-\dif U(\tilde{X})U^\dagger.
\end{align}
Due to the arbitrariness of $\tilde{X}$, we obtain Eq.~\eqref{AU9}.

Again, we assume $\tilde{X}$ is the tangent vector of the horizontal curve $\tilde{\gamma}$ parameterized by $t$, then $\tilde{X}=\frac{\dif}{\dif t}$. Thus, Eq.~(\ref{AU8}) now reads
\begin{align}\label{AU8'}
A_U(X)=-\frac{\dif U}{\dif t} U^\dagger,
\end{align}
or
\begin{align}\label{AU8''}
U^\dagger A_U(X) U+U^\dagger \frac{\dif U}{\dif t} =0.
\end{align}
Eq.~(\ref{AU8''}) is equivalent to Eq.~(\ref{OmegaXU}) when $\omega$ is given by Eq.~(\ref{AU4}), which is also a generalization of Eq.~(\ref{OmegaX1}). Now integrating Eq.~(\ref{AU8'}) along the loop $\gamma=\pi\circ\tilde{\gamma}$ from $0$ to $1$, we get
\begin{align}\label{W8}U(1)=\mathcal{P}\me^{-\int_0^1A_U(X)\dif t}U(0)=\mathcal{P}\me^{-\oint A_U}U(0).\end{align}

To derive the expression of $A_U$, we note that Eq.~(\ref{AU6}) can also be written as
\begin{align}\label{AU6'}
\rho A_U+A_U\rho=-[\dif\sqrt{\rho},\sqrt{\rho}].
\end{align}
Again, d is the ``horizontal lift'' of the exterior derivative on $Q$. Taking the matrix elements of both sides of Eq.~(\ref{AU6'}) with respect to the eigenvectors of $\rho$, we have
\begin{align}\label{AU61}
(\lambda_i+\lambda_j) \langle i|A_U|j\rangle=-\langle i|[\dif\sqrt{\rho},\sqrt{\rho}]|j\rangle.
\end{align}
Therefore, we obtain Eq.~\eqref{AUE}.
Finally, the Uhlmann phase is given by
\begin{align}
\theta_U & =\arg\langle W(0)|W(1)\rangle=\arg\textrm{Tr}[W(0)^\dagger W(1)] \notag \\
&=\arg\textrm{Tr}[\rho(0)\mathcal{P}\me^{-\oint A_U}].%\notag\\
%&=\arg\textrm{Tr}[W(0)^\dagger W(1)]\notag\\
%&=\arg\textrm{Tr}[U^\dagger(0)\sqrt{\rho(0)}\sqrt{\rho(0)}U(1)]\notag\\
%&=\arg\textrm{Tr}[U^\dagger(0)\sqrt{\rho(0)}\sqrt{\rho(0)}U(1)]\notag\\
%&=\arg\textrm{Tr}[\rho(0)U(1)U^\dagger(0)]\notag\\
%&=\arg\textrm{Tr}[\rho(0)\mathcal{P}\me^{-\oint A_U}],
\end{align}
%\textbf{Please give an equation for the definition of Uhlmann phase.}

\subsection{Parallel-Transport Condition}\label{appUhlmann4}
Eq.~(\ref{W8}) gives the parallel condition between two phase factors when $\rho(1)=\rho(0)$. A more general parallel condition also exists even if $\rho(1)\neq\rho(0)$. %$\rho_1\neq \rho_2$. %\textbf{Did you mean $\rho(1)\neq\rho(0)$?}
Note the right-hand-side of Eq.~(\ref{UC2}) vanishes on any horizontal vector field. At any horizontal direction, we have
\begin{align}\label{upcd5}
W^\dagger \dif W=\dif W^\dagger W.
\end{align}
Integrating both sides along a horizontal curve, we obtain Eq.~\eqref{upcd6}.
This is a generalization of Eq.~(\ref{p1}).
Plugging in $W_1=\sqrt{\rho_1}U_1$ and $W_2=\sqrt{\rho_2}U_2$, we get
\begin{align}\label{rr1}
\sqrt{\rho_2}\sqrt{\rho_1}=U_2U^\dagger_1\sqrt{\rho_1}\sqrt{\rho_2}U_2U^\dagger_1.
\end{align}
Multiplying both sides by their complex conjugates from the left, we have
\begin{align}\label{rr2}
\sqrt{\rho_1}\rho_2\sqrt{\rho_1}&=(U_2U_1^\dagger)^\dagger\sqrt{\rho_2}\rho_1\sqrt{\rho_2}U_2U_1^\dagger \notag \\
&=(\sqrt{\rho_1}\sqrt{\rho_2}U_2U_1^\dagger)^\dagger\sqrt{\rho_1}\sqrt{\rho_2}U_2U_1^\dagger.
\end{align}
Eq.~(\ref{upcd6}) indicates that $\sqrt{\rho_1}\sqrt{\rho_2}U_2U_1^\dagger$ is Hermitian.
Therefore, we finally get
\begin{align}\label{UU0}
U_2U^\dagger_1=\sqrt{\rho_2^{-1}}\sqrt{\rho_1^{-1}}\sqrt{\sqrt{\rho_1}\rho_2\sqrt{\rho_1}}.
\end{align}

The polar decomposition theorem tells us that any full-ranked matrix $A$ can be decomposed as $A=|A|U_A$, where $|A|=\sqrt{AA^\dagger}$ and $U_A$ is a unitary matrix. For an arbitrary unitary matrix $U$, the follow inequality holds~\cite{2DMat15}
\begin{align}\label{ieq}
\textrm{Re}\big[\textrm{Tr}(AU)\big]\leq \textrm{Tr}|A|.
\end{align}
It
is saturated if $\sqrt{|A|}=\sqrt{|A|}U_AU$, i.e. $U=U^\dagger_A$.
Applying this to Eq.~(\ref{d1}), we have
\begin{align}\label{d2a}
\textrm{Re}\big[\textrm{Tr}(W^\dagger_1W_2)\big]&=\textrm{Re}\big[\textrm{Tr}(U^\dagger_1\sqrt{\rho_1}\sqrt{\rho_2}U_2)\big]\notag\\
%&=\textrm{Re}\big[\textrm{Tr}(\sqrt{\rho_1}\sqrt{\rho_2}U_2U^\dagger_1)\big]\notag\\
&\leq \textrm{Tr}|\sqrt{\rho_1}\sqrt{\rho_2}|\notag\\
&=\textrm{Tr}\sqrt{\sqrt{\rho_1}\rho_2\sqrt{\rho_1}},
\end{align}
where the last line is called the fidelity of the two density matrices.
Therefore, Eq.~(\ref{d1}) is satisfied in the generic situation, and the equal sign is satisfied when
\begin{align}
U_2U^\dagger_1&=U^\dagger_{\sqrt{\rho_1}\sqrt{\rho_2}}=(|\sqrt{\rho_1}\sqrt{\rho_2}|^{-1}\sqrt{\rho_1}\sqrt{\rho_2})^{-1} \notag \\
&=\sqrt{\rho_2^{-1}}\sqrt{\rho_1^{-1}}\sqrt{\sqrt{\rho_1}\rho_2\sqrt{\rho_1}},
\end{align}
which agrees with Eq.~(\ref{UU0}).

In fact, the parallel-transport condition~\eqref{upcd6} is also the condition that minimizes the Hilbert-Schmidt distance between two amplitudes:
\begin{align}\label{d1}
\dif^2_\text{HS} (W_1,W_2)&=\inf\textrm{Tr}(W_1-W_2)^\dagger(W_1-W_2) \notag \\
&=2-2\sup\textrm{Re}\big[\textrm{Tr}(W_1^\dagger W_2)\big].
\end{align}
which is a natural metric associated with the scalar product defined by Eq.~(\ref{HSip}). Therefore, the distance between two amplitudes is minimized if the parallel transport condition is satisfied, and the distance (\ref{d1}) is now given by
\begin{align}\label{E2a}
\dif^2_\text{HS}(W_1,W_2)=2-2\textrm{Tr}\sqrt{\sqrt{\rho_1}\rho_2\sqrt{\rho_1}},
\end{align}
which is the Bures distance between the density matrices $\rho_1$ and $\rho_2$.

%\bibliographystyle{apsrev}
%\bibliography{Review,Review1}

\end{document}